\documentclass[preprint,12pt]{elsarticle}

\usepackage{amssymb}
\usepackage{amsmath}

\journal{Journal of magnetic resonance}
\usepackage{booktabs}
\usepackage{algorithm}
\usepackage{algpseudocode}
\usepackage{hyperref}
 \usepackage[margin=1in]{geometry}

\begin{document}

\begin{frontmatter}


\title{Input layer regularization and automated regularization hyperparameter tuning for myelin water estimation using deep learning.}
\author[1]{Mirage Modi}{} 
\author[2]{Shashank Sule}{} 
\author[1]{Jonathan Palumbo}{} 
\author[2]{Michael Rozowski}{}
\author[1]{Mustapha Bouhrara}{}
\author[2]{Wojciech Czaja}{}
\author[1]{Richard G. Spencer}{}

\affiliation[1]{organization={National Institute on Aging, National Institutes of Health}}
\affiliation[2]{organization={Department of Mathematics, University of Maryland, College Park}}




\begin{abstract}
    We propose a novel deep learning method which combines classical regularization with data augmentation for estimating myelin water fraction (MWF) in the brain via biexponential analysis. Our aim is to design an accurate deep learning technique for analysis of signals arising in magnetic resonance relaxometry. In particular, we study the biexponential model, one of the signal models used for MWF estimation. We greatly extend our previous work on \emph{input layer regularization (ILR)} in several ways. We now incorporate optimal regularization parameter selection via a dedicated neural network or generalized cross validation (GCV) on a signal-by-signal, or pixel-by-pixel, basis to form the augmented input signal, and now incorporate estimation of MWF, rather than just exponential time constants, into the analysis. On synthetically generated data, our proposed deep learning architecture outperformed both classical methods and a conventional multi-layer perceptron. On in vivo brain data, our architecture again outperformed other comparison methods, with GCV proving to be somewhat superior to a NN for regularization parameter selection. Thus, ILR improves estimation of MWF within the biexponential model. In addition, classical methods such as GCV may be combined with deep learning to optimize MWF imaging in the human brain.
\end{abstract}








\end{frontmatter}



\section{Introduction}\label{sec: Background}
\subsection{Myelin water fraction imaging}
\label{subsec: MyelinIntro}

Myelin is a lipid-rich substance that surrounds and insulates nerve axons in the central and peripheral nervous systems, facilitating the rapid and high-fidelity transmission of electrical signals. Therefore, quantification and mapping of myelin provides valuable insights into brain white matter. In addition, myelination in the central nervous system (CNS) is increasingly recognized as being of central importance in a spectrum of neurodegenerative disorders, including Alzheimer's disease and cognitive impairment \citep{faizy2018age}, making it an important biomarker for neurodegenerative disorders such as Alzheimer's disease, multiple sclerosis (MS), and traumatic brain injury (TBI) \cite{laule2004water, dula2010multiexponential, mackay1994vivo, kolind2012myelin, mackay2016magnetic}. Consequently, the problem of quantification and mapping of the myelin water fraction (MWF) using MRI remains a very active area of investigation, even after its introduction through the pioneering work of MacKay more than 20 years ago \citep{mackay1994vivo}. 

There are a number of current methods for MWF mapping; relaxation-based methods rely upon use of an MRI signal model that incorporates T2 values of both myelin-associated water and less-restricted water. In this paper we address multiple gradient-echo or spin-echo pulse sequences \citep{vavasour2006magnetization, deoni2010quantitative,oh2007multi} which result in the classic inverse problem of biexponential parameter estimation \citep{istratov1999exponential} of substantial independent mathematical interest. In the case of MWF mapping, the method relies upon the more rapid transverse relaxation of the less-mobile water entrapped within myelin lamallae as compared to the slower rate of transverse relaxation of more-mobile water. In effect, the problem is that of separating the two exponential decay signals that comprise the net acquired signal. Depending primarily upon the difference in the decay constants of the two underlying signals, this can be an extremely challenging parameter estimation problem \citep{landaw1984multiexponential}. 

An alternative approach is based on the regularized NNLS approach, sometimes described as model-independent because there is no assumption regarding the number of exponentially decaying components \citep{mackay1994vivo}. While it can be shown that the two approaches produce very similar results for MWF, our purpose here is not to compare the two methods, but to develop ILR, suitable for the approach for which the model is a biexponential decaying signal.

The outline of the paper is as follows, which also serves to collect several symbols and definitions used. In Section~\ref{sec:math_formalism}, describing general features of the problem set up, we first define the biexponential model with Rician noise, and outline the NLLS approach to parameter estimation in this model as an inverse problem. We note the requirement for regularization and outline methods for regularization parameter selection, including a $\lambda$ selection NN, denoted $\lambda$-NN. Finally, we describe NN input layer regularization (ILR), the data augmentation method we present here for improved parameter estimation. 

In Section~\ref{sec:methods}, we describe two methods of ILR implementation, based respectively on selection of $\lambda$ via generalized cross validation (GCV) or via the $\lambda$-NN, denoted respectively by $\lambda_{GCV}$ and $\lambda_{NN}$. In both cases, the selection of $\lambda$ is followed by the implementation of the parameter-estimation (PE) NN using ILR. This involves concatenating the noisy decay (ND) vector with a version of itself regularized with either $\lambda_{GCV}$ or $\lambda_{NN}$ to form the input signal to the PE NN, with the concatenated input signals denoted $\textsc{(ND, Reg)}_{\textsc{GCV}}$ or $\textsc{(ND, Reg)}_{\textsc{NN}}$.  Since this involves doubling the length of the input ND signal, we compare performance of the two ILR methods with a PE NN for which the input vector is the noisy signal concatenated with itself, deemed $\textsc{(ND, ND)}$.

 Although our primary interest is in parameter estimation, the performance of the $\lambda$-NN is of independent interest, given ongoing  research in $\lambda$ selection for Tikhonov regularization. Therefore, in Section~\ref{sec:results}, we use an oracle $\lambda$ as a comparison standard for evaluating the performance of GCV and the $\lambda$-NN for regularizer selection on simulated data. We then compare five methods of parameter estimation, focusing on MWF; these are three NN's, with subscripts indicating the method of $\lambda$ selection: 
 \begin{enumerate}
     \item $\textsc{(ND, Reg)}_{\textsc{NN}}$, 
\item $\textsc{(ND, Reg)}_{\textsc{GCV}}$, 
\item $\textsc{(ND, ND)}$ as described above; 
\item Tikhonov-regularized NLLS (TR-NLLS), using $\lambda_{GCV}$, and 
\item conventional NLLS.
 \end{enumerate}
 These are all compared to TR-NLLS using the optimal, "oracle", value of $\lambda$. We then present results for in vivo human brain. In this case, there is no true oracle standard of comparison, so we establish a high-quality MWF map as a reference standard. Of note is that we apply ILR only to pixels for which the decay signal is of biexponential character. We compare the performance of the two ILR PE NN's, using respectively $\lambda_{NN}$ and $\lambda_{GCV}$, finding that the latter exhibits somewhat superior performance, and then compare this with the conventional $\textsc{(ND, ND)}$ PE NN, not using ILR. 

In Section~\ref{sec:discussion}, we highlight additional aspects of the underlying problem and certain limitations of our approach.

\section{Mathematical formalism}
\label{sec:math_formalism}

\subsection{Biexponential signal analysis}

The noiseless biexponential model \citep{istratov1999exponential} for spin-echo imaging is of the form:
\begin{equation} \label{eq:1}
s(t; T_{2,1}, T_{2,2}, c_1, c_2) = c_1e^{\frac{-t}{T_{2,1}}} + c_2e^{\frac{-t}{T_{2,2}}},
\end{equation}
with $s(t)$ indicating signal amplitude at echo time $t$. The estimation problem is to extract the component amplitudes $c_1$ and $c_2$, and the corresponding spin-spin decay constants $T_{2,1}$ and $T_{2,1}$ \citep{ansorge2016physics} \citep{alonso2015mri, mackay2016magnetic}. 

The model is linear with respect to $c_1,c_2 > 0$ but nonlinear with respect to corresponding decay constants $T_{2,1}, T_{2,2}$.These two components represent two distinct water compartments in tissue. In the present work, we restrict our analysis to the three-parameter problem in which $c_2 = 1 - c_1$. It is assumed that $s(t)$ is measured at a finite number of echo times $\{t_n\}_{n=1}^N$; Eq. \eqref{eq:1} becomes: 
\begin{equation} \label{eq:Biexp}
\textbf{G(p)} \equiv (c_1e^{\frac{-t_n}{T_{2,1}}} + (1-c_1)e^{\frac{-t_n}{T_{2,2}}})^N_{n=1},
\end{equation}
where $\textbf{p}$ is the set of parameters $(c_1,T_{2,1}, T_{2,2})$.
We will further assume $t_1 = 0$ as is usual for biexponential analysis, although for the conventional MRI experiment data acquisition is initiated at a minimum echo time $TE_0 >0$ which depends primarily on gradient and pulse duration.  Additionally, \textbf{G(p)} is invariably corrupted by noise, leading to the general signal model: 
\begin{equation} \label{eq:noisy signal}
\textbf{s} = \mathcal{M}(\textbf{G(p)})
\end{equation}
 where $\mathcal{M}$ denotes the noisy measurement process. In this paper we will address the conventional case of MR magnitude imaging, in which the noise model is Rician so that 
 \begin{align}
     \textbf{s} = \sqrt{(\textbf{G(p)} + \xi)^2 + \eta^2} \label{eq: rician noise model}
 \end{align}
 Here $\xi$ and $\eta$ are Gaussian random variables of zero mean and equal variance. When the signal \textbf{s} is normalized to unity, the corresponding  signal-to-noise ratio (SNR) will be expressed as the standard deviation $\sigma$ of these random variables. In other applications, zero-mean additive Gaussian random noise will be the most common noise model. We denote the true underlying parameter set producing the signal $\textbf{G}$ as $\textbf{p}_{true}$. The goal of the analysis is to recover the parameter vector $\textbf{p}_{true}$ from the noisy measurement vector $\textbf{s}$.

\subsection{Parameter Estimation}
\label{subsec: PEMIntro}

Although our interest is in the comparison of different NN architectures for biexponential analysis, we include a non-linear least squares (NLLS) benchmark as well. We note that for Rician data, this deviates from statistically validated MLE analysis \citep{sijbers1998maximum,jiang2003adaptive}, but is nevertheless conventional and accurate for reasonably high SNR. The NLLS approach involves computing an estimate $\textbf{p}^*$ that minimizes the mean-squared difference between the model $\textbf{G(p)}$ and the signal $\textbf{s}$ for $\textbf{p} \in F$, where $F$ is a feasible set of parameters. We can tailor $F$ based on our problem, in this case constraining $T_{2,1}, T_{2,2} \geq 0$ and $0 \leq c_1 \leq 1$. Therefore the NLLS approach to parameter estimation is mathematically formulated as follows: 
\begin{align}\label{eq:3}    
        \textbf{p}^* &\equiv \textsf{argmin}_{\textbf{p} \in F}{\;\lVert{\textbf{G(p)} - \textbf{s}}\rVert}^2_2 \\
        &= \!\!\!\!\!\!\!\!\!\! \underset{T_{2,1}, T_{2,2} \geq 0, 0 \leq c_1 \leq 1}{\textsf{argmin}}\left[\sum_{n=1}^{N} (c_1e^{\frac{-t_n}{T_{2,1}}} + (1-c_1)e^{\frac{-t_n}{T_{2,2}}} - s_n)^2 \right].
\end{align}
However, as stated, the inverse problem in Eq. \eqref{eq:3} is highly ill-posed in the sense of Hadamard \citep{aster2018parameter}, meaning that distinct parameter choices $\mathbf{p_1}$ and $\mathbf{p_2}$ can map to nearly identical biexponential decay curves \citep{lanczos1988applied}. This ill-posedness results in a high sensitivity to noise of derived parameter estimates. Stabilization of the parameter estimation problem is therefore a major topic of investigation in inverse problems.

\subsection{Regularization}

\label{subsec: RegIntro}
A well-known approach to address the ill-posedness of the problem Eq. \eqref{eq:3} is \emph{regularization}, in which Eq. \eqref{eq:3} is modified through the introduction of an additional term to the objective function that penalizes the complexity of feasible solutions $\textbf{p}$ \citep{calvetti2018inverse}. Among the many ways of quantifying complexity, an effective approach is to supplement Eq. \eqref{eq:3} with \emph{Tikhonov} regularization (TR) \citep{golub1999tikhonov}, where the complexity is described through the $L^2$ norm of the parameters weighted against the fidelity term through the \emph{hyperparameter} $\lambda$: 
\begin{align}
    \textbf{p}^{*}_{\lambda(\textbf{s})} := \underset{\textbf{p} \in F}{\textsf{argmin}} \|\textbf{G(p)} - \textbf{s}\|^{2}_{2} + \lambda \|\textbf{p}\|^2.\label{eq:RNLLS}
\end{align}

Along with other variational regularization techniques such as Lasso ($L^1$ regularization) or elastic net (a linear combination of both $L^2$ and $L^1$ regularizers), TR effectively serves to stabilize ill-posed inverse problems. In particular, the parameter $\lambda \geq 0$ improves the condition number of the source-to-solution map that takes the noisy signal $\textbf{s}$ to the parameter $ \textbf{p}_\lambda^*(\textbf{s})$, henceforth termed the \emph{Tikhonov Regularized NLLS (TR-NLLS)} estimate. In a Bayesian framework, it provides Gaussian priors on the parameters where the TR-NLLS estimate is the maximum a posteriori (MAP) estimator. In the present case, there is no straightforward statistical or physical interpretation of the TR-NLLS estimator $\textbf{p}_{\lambda}^{*}(\textbf{s})$ and the application of TR for parameter estimation in Rician-distributed signals is highly non-standard. However, even lacking a true physical basis, regularization applied in Eq. \eqref{eq:RNLLS} TR-NLLS would be expected to improve stability. 

\subsection{Regularization Parameter Selection}

\label{subsect:LambdaSelectionIntro}
The appropriate selection of $\lambda$, which controls the bias-variance tradeoff in estimating $\textbf{p}_{true}$ through TR, is a central challenge in inverse problems. Large values lead to simple, low complexity solutions (low variance) but with low fidelity to the underlying signal $\textbf{s}$ (high bias) and highly biased parameter estimates. Conversely, small values ensure the closest fidelity to the signal (low bias) but high variability to small perturbations in the signal (high variance), such as due to noise. In general, the problem of choosing the optimal $\lambda$ refers to selecting the $\lambda$ for which the associated solution to Eq. \eqref{eq:RNLLS} $\textbf{p}_{\lambda}^{*}(s)$ is the closest to the true underlying parameter $\textbf{p}_{true}$. This  can be mathematically formulated through the following \emph{bilevel} optimization problem: 
\begin{align}
    \lambda_{oracle}(\textbf{s}) &= \underset{\lambda}{\textsf{argmin}} \|\textbf{p}_{true} - \textbf{p}_{\lambda}^{*}(\textbf{s}) \|, \quad \text{w.r.t} \label{eq: upper level}\\
    \textbf{p}_\lambda^*(\textbf{s}) = & \underset{\textbf{p} \in F}{\textsf{argmin}} {\lVert{\textbf{G(p)} - \textbf{s}}\rVert}^2_2 + \lambda\lVert{\textbf{p}}\rVert_2^2.  \label{eq: lower level}
 \end{align}
Here Eq. \eqref{eq: upper level} is referred to as the \emph{upper level} problem and Eq. \eqref{eq: lower level} is referred to as the \emph{lower level} problem. Note that the solution to the upper level problem depends upon the noisy signal $\textbf{s}$ which in turn depends on the map $\textbf{G}$ and the SNR. Thus the bilevel formulation identifies two sources which affect the optimal $\lambda$. We term the optimal $\lambda(\textbf{s})$ from Eq. \eqref{eq: upper level} as $\lambda_{oracle}$ and refer to the following problem as the $\lambda-$\emph{selection problem}: 
\begin{align}
    \text{Recover } \lambda_{oracle} \text{ from } \textbf{s} \label{eq: lambda selection problem}
\end{align}
Typically it is infeasible to directly solve the bilevel problem Eq. \eqref{eq: upper level}-\eqref{eq: lower level} because $\textbf{p}_{true}$ is not available in applications. Moreover, the bilevel problem Eq. \eqref{eq: upper level}-\eqref{eq: lower level} is difficult to solve, and theoretical guarantees for the existence and positivity of solutions are an active area of research \citep{holler2018bilevel}. Consequently, a great deal of theory is dedicated to approximating $\lambda_{oracle}$ using alternative formulations and algorithmic techniques\citep{alberti2021learning, theodoridis2015machine}. In practice, numerical methods for choosing $\lambda$ are often implemented, including the $L$-curve (LC) \citep{engl1994using}, discrepancy principle (MDP) \citep{nair2003morozov}, or generalized cross validation (GCV) \citep{golub1979generalized}. Among these, GCV stands out as a theoretically and practically robust method for approximating $\lambda_{oracle}$ \citep{lukas1993asymptotic, golub1997generalized, haber2000gcv}. 

\subsection{Neural networks for parameter estimation}
An alternative approach to $\lambda$-selection is through the use of a neural network (NN) to design a map from the noisy signal $\textbf{s}$ to an optimal $\lambda$  \citep{Afkham, kobler2020total,li2020nett}. A fully connected $L$-layer NN is defined by the alternating composition of linearities and nonlinearities: 
\begin{align}
    f_{\Theta}(x) = W_L\left(\sigma(W_{L-1}(\ldots x)) + b_{L-1}\right) + b_{L} \label{eq: nn}
\end{align}
Given a training set consisting of input-output pairs $\{x_i, y_i\}_{i=1}^{m}$ (such as pairs of noisy signals and their corresponding biexponential parameters $\{\textbf{s}_i, \textbf{p}^{i}_{true}\}$), the network weights, $W_i$, and biases, $b_i$, where we write $\Theta = \{W_i, b_i\}_{i=1}^{L}$,  are determined by optimizing a loss function $\mathcal{L}$ measuring the discrepancy between the true outputs $y_i$ and the NN outputs $f(x_i)$: 
\begin{align}
    \Theta = \underset{\theta}{\textsf{argmin}}\frac{1}{m}\sum_{i=1}^{m}\mathcal{L}(y_i, f_{\theta}(x_i)) \label{eq: nn training}
\end{align}
NNs are well-established in the setting of inverse problems \citep{ching2018opportunities} in the biomedical sciences and offer a model-free approach to mapping an unseen signal to parameter estimates. Unlike classical mapping methods (such as TR-NLLS), NNs require training with synthetically or experimentally generated data. Such approaches involving NNs routinely outperform classical methods as long as the underlying NN is well-trained\citep{saniei2016parameter,worswick2018deep,almeida2003neural}. In fact, NNs and classical methods like TR-NLLS may be combined: in \citep{Afkham}, a parameter estimation strategy was developed where an estimate for $\textbf{p}_{true}$ is obtained through TR-NLLS where the $\lambda$ in Eq. \eqref{eq:RNLLS} is chosen using a neural network. These considerations allow us to define two distinct NNs for the present problem: 
\begin{enumerate}
    \item NNs for $\lambda$-selection: Following \citep{Afkham}, we develop a NN for $\lambda$ selection (termed $\lambda_{NN})$ trained with the input-output pairs $\{\textbf{s}_i, \lambda_{oracle}(\textbf{s}_i)\}$. We compare this with other the methods for choosing $\lambda$ mentioned in Section \ref{subsect:LambdaSelectionIntro}. 
    \item NNs for parameter estimation: We develop a separate NN for estimations of the parameter $\textbf{p}_{true}$ from the noisy signal $\textbf{s}_i$. 
\end{enumerate}

\subsection{Input Layer Regularization}

\begin{figure}[h]
    \centering
    \includegraphics[width=\textwidth]{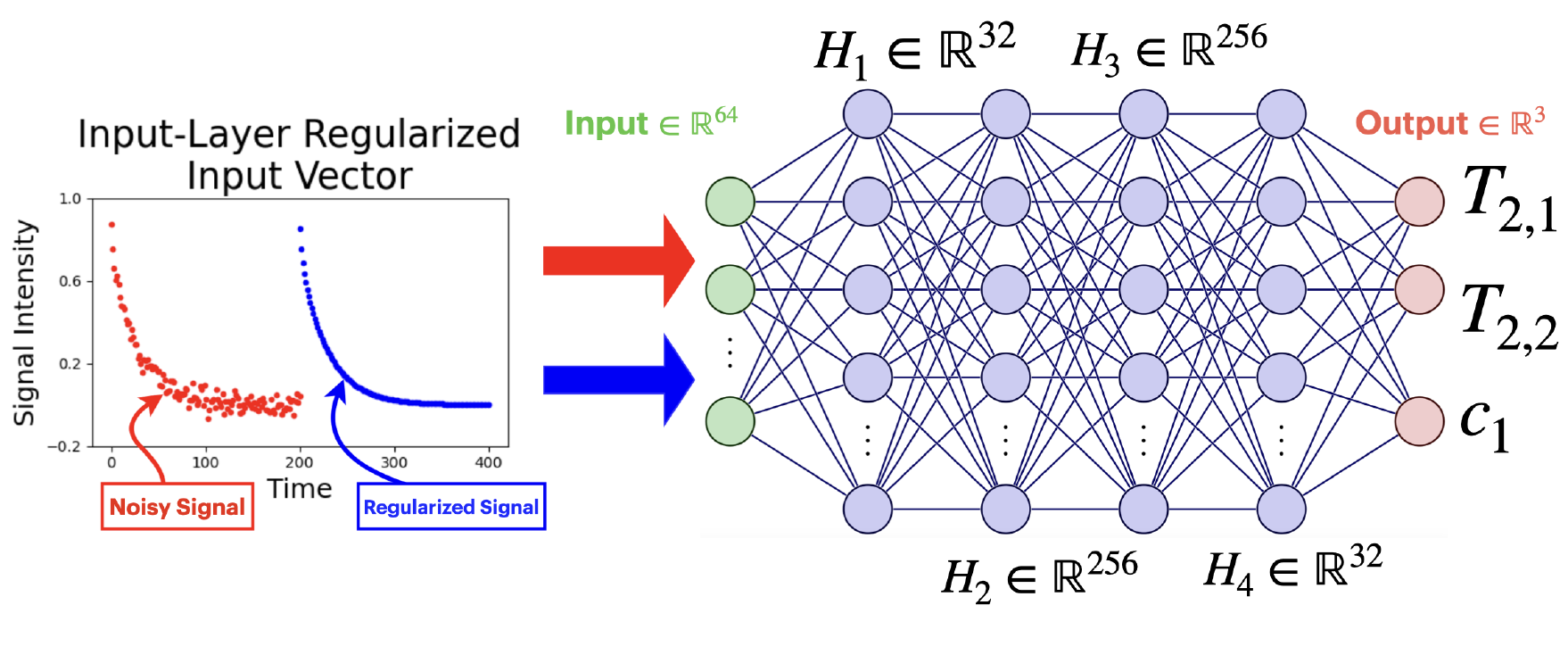}
    \caption{The (ND, Reg) parameter estimation network is an NN where the inputs are the concatenated vectors of noisy signal $s$ and the signal obtained from parameters estimated by Tikhonov regularization, with regularization parameter $\lambda(\textbf{s})$ given by either a neural network or GCV. The parameters estimated are  $c_1, T_{2,1}$ and $T_{2,2}$.}
    \label{fig:peNN}
\end{figure}

In this paper we combine $\lambda$-selection, via a NN or classical techniques, with a NN for parameter estimation into a pipeline for estimating $\textbf{p}_{true}$. The result is a data augmentation technique that we call \emph{input layer regularization (ILR)}\citep{rozowski2022input}. In ILR, the noisy data (ND) $\textbf{s}$ is concatenated with, or augmented by, a regularized (Reg) version given by $G\left(\textbf{p}^{*}_{\lambda(\textbf{s})}\right)$ where $\textbf{p}^{*}_{\lambda(\textbf{s})}$ is a TR-NLLS parameter estimate from Eq. \eqref{eq:RNLLS} with hyperparameter $\lambda(\textbf{s})$ chosen through a user-defined $\lambda$-selection method. The augmented input vector is thus given by
\begin{align}
    \textbf{x} := \left[\textbf{s}, \textbf{G}(\textbf{p}^{*}_{\lambda(\textbf{s})}(\textbf{s}))\right]. \label{eq: concatenated vector}
\end{align}
A NN (termed $\textsc{(ND, Reg)}$) is trained to estimate $\textbf{p}_{true}$ from $\textbf{x}$ (see Figure \ref{fig:peNN}). ILR can thus be viewed as an unconventional data augmentation strategy. Besides our unconventional application here, data augmentation has been established as a proven means for \emph{implicit regularization} of a parameter estimation NN. In particular, it has been shown in image classification settings that data augmentation with invariant transformations (such as rotations, translations, and scaling) prevents overfitting and enables the NN to generalize more accurately to unseen data \citep{shorten2019survey}. It is a priori unclear how to transplant these data augmentation techniques from image classification to the context of MRI because the noisy signals $\textbf{s}$ have no clear invariances. The central innovation of ILR is to introduce the smooth vector $G\left(\textbf{p}^{*}_{\lambda(\textbf{s})}\right)$ as for augmentation of the raw input data. ILR was shown to improve the two-parameter biexponential estimation problem by $\sim10 \%$ when compared to (1) NLLS and (2) NN estimation without augmentation of the input \citep{rozowski2022input}. However, substantial limitations in \citep{rozowski2022input} include (1) the very limited application to the case where the MWF $c_1$ is known,  and (2) use of a single pre-defined $\lambda$ for construction of the augmented component of the input signal vector for every noisy decay signal $\textbf{s}$. Here, we greatly extend the analysis by solving the 3 parameter problem, permitting an estimate of MWF, and incorporating signal-dependent $\lambda$-selection via NNs and GCV.  
   
\section{Methods} 
\label{sec:methods}
\begin{figure*}
    \centering
    \includegraphics[width=0.9\textwidth]{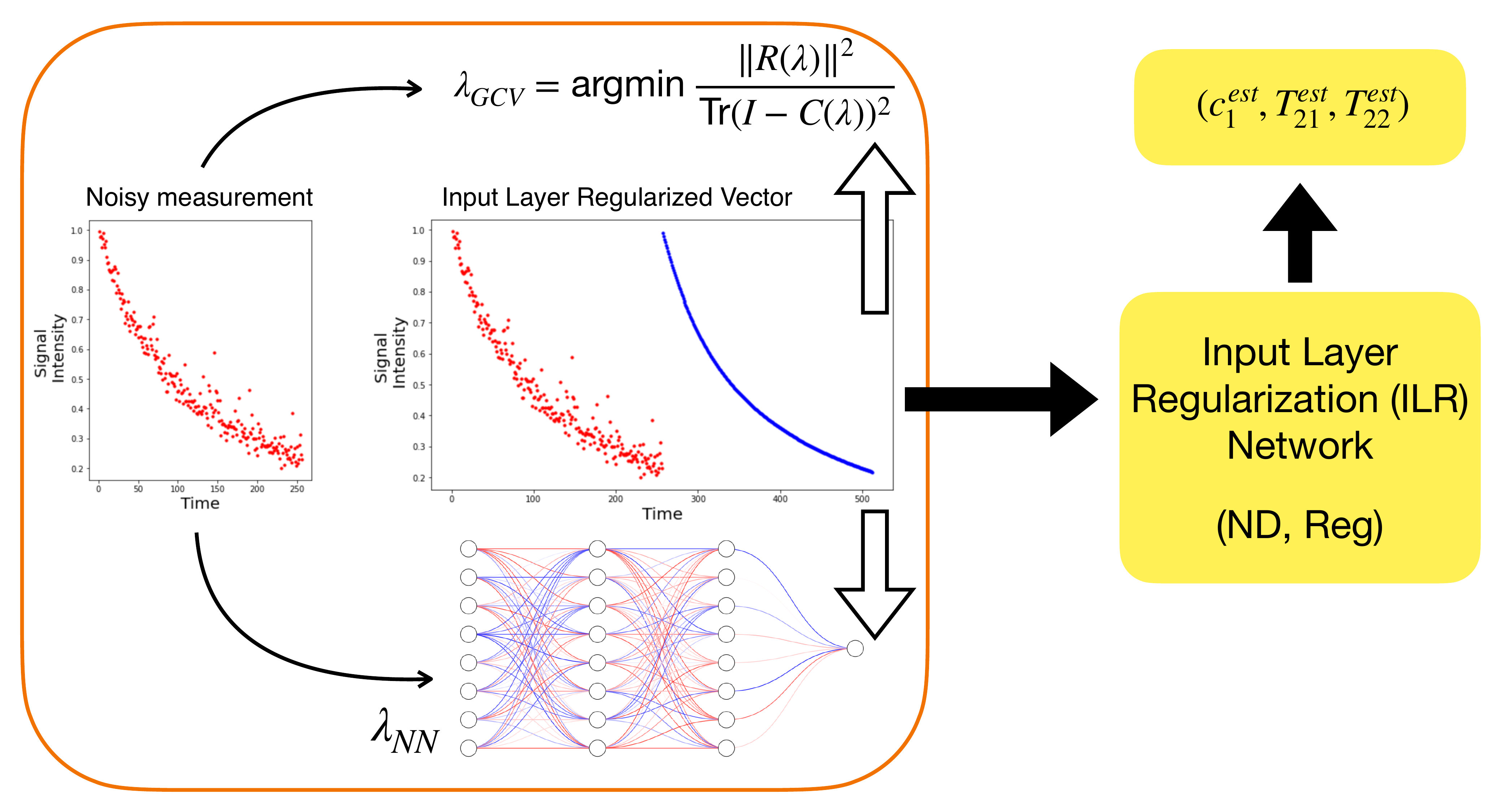}
    \caption{Combination of signal-dependent $\lambda$ selection with input layer regularization. The first step involves selecting either a NN or GCV for estimating $\lambda$. Training data for the ILR networks $\textsc{(ND, Reg)}$ are first passed through a TR-NLLS solver with the signal-dependent $\lambda(\textbf{s})$ to construct the concatenated vector $\textbf{x}$ in Eq. \eqref{eq: concatenated vector}. $C(\lambda)$ is defined according to Eq.  \eqref{eq: GCV function}}
    \label{fig: overall workflow}
\end{figure*}

As above, we propose a methodology combining ILR with signal-based selection of $\lambda$ for solving the parameter estimation problem. We compare two choices for generating $\lambda$: (1) a neural network, following \citep{Afkham}) and (2) GCV (outlined in Section \ref{subsec: gcv}). See Figure \ref{fig: overall workflow}. In Section \ref{subsect: param estimation}, we discuss the choice between use of $\lambda_{NN}$ and $\lambda_{GCV}$. 

We now outline our algorithm for each choice of estimating $\lambda$. 

\subsection{Selecting $\lambda$ with a $NN$ followed by parameter estimation}

\begin{figure*}
    \centering
    \includegraphics[width=0.8\textwidth]{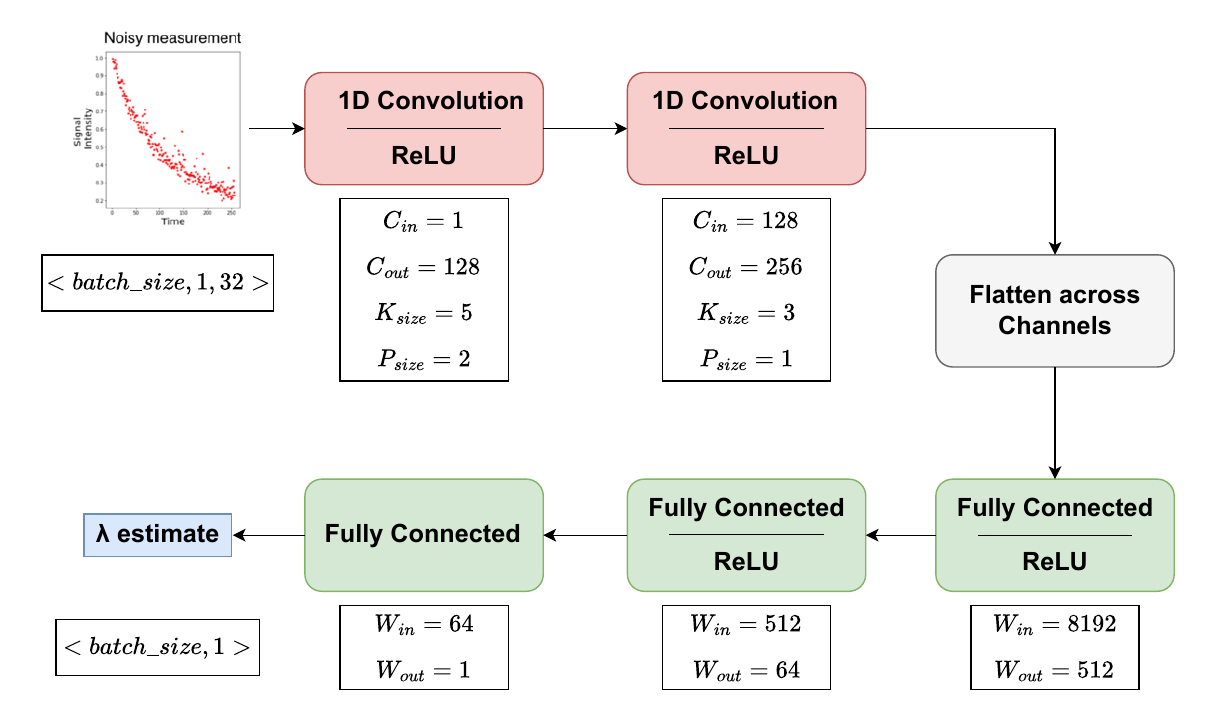}
    \caption{$\lambda_{NN}$ architecture. 1-dimensional convolutional layers are applied to the input noisy signal, followed by fully connected layers to predict $\lambda_{oracle}$. $C$ refers to the channel sizes (in/out for input or output channels, respectively). We also define the kernel size as $K_{size}$ and the padding as $P_{size}$. $W$ refers to the width of a given fully-connected layer. The ReLU activation function is used between each layer.}
    \label{fig:lamArchitecture}
\end{figure*}

To estimate $\lambda$ by a NN, $\lambda(\textbf{s}) := \lambda_{NN}(\textbf{s})$, we train $\lambda_{NN}$ using synthetically generated training data $\{\textbf{s}_i, \lambda_{oracle}(\textbf{s}_i)\}$. We summarize the data generation process in Section \ref{subsec: datagen}. Here $\lambda_{NN}$ is a convolutional neural network (CNN) with the architecture shown in Figure \ref{fig:lamArchitecture}. We make use of $L^1$ loss to train the NN to recognize $\lambda_{oracle}$ in order not to heavily penalize outliers since the distribution of $\lambda_{oracle}$ spans several orders of magnitude (Figure \ref{fig:NNvsGCV_distribution}). Thus, the loss function used for training $\lambda_{NN}$ is given by: 
\begin{align}
    \mathcal{L}_{\lambda} := \frac{1}{N}\sum_{i=1}^{N}|\lambda_{oracle}(\textbf{s}_i) - \lambda_{NN}(\textbf{s}_i)| \label{eq: lambda NN loss function}
\end{align}
Our approach is summarized in Algorithm \ref{alg:Lambda-Selection NN}.

To recover parameter estimates, we generate the concatenated input $\textbf{x}_i$ according to Eq. \eqref{eq: concatenated vector} with $\lambda_{NN}(\textbf{s})$ in place of $\lambda(\textbf{s.})$. Letting $\textbf{p}_i$ be the parameter vector used for generating the noisy signal $\textbf{s}_i$, we train a neural network $\textsc{(ND, Reg)}_{\textsc{NN}}$ for recovering $\textbf{p}_{i}$ from $\textbf{x}_i$ by minimizing mean-squared error. The loss function for training $\textsc{(ND, Reg)}_{\textsc{NN}}$ is given by: 
\begin{align}
    \mathcal{L} := \frac{1}{N}\sum_{i=1}^{N}\|\textbf{p}_{i} -\textsc{(ND, Reg)}_{\textsc{NN}}\left(\textbf{x}_i\right) \|_{2}^{2} \label{eq: NDReg NN loss function}
\end{align}

\subsection{Selecting $\lambda$ with GCV followed by parameter estimation}
\label{subsec: gcv}
Here we describe the algorithm used for producing $\lambda_{GCV}(\textbf{s})$. Recall that $\textbf{G}$ in Eq. \eqref{eq:Biexp} maps a parameter combination $\textbf{p} \in \mathbb{R}^{3}$ to a noiseless signal vector $\textbf{G}(\textbf{p}) \in \mathbb{R}^{N}$. $\textbf{G}$ is a differentiable map from $\mathbb{R}^3$ to $\mathbb{R}^{N}$ and admits a Jacobian $D\textbf{G}: \mathbb{R}^{3} \to \mathbb{R}^{N}$. Next, recall that for a fixed $\lambda$, $\textbf{p}^{*}_{\lambda}(\textbf{s})$ is the TR-NLLS parameter estimate Eq. \eqref{eq:RNLLS}. This allows us to define a Jacobian $D\textbf{G}$, a matrix-valued function that depends \emph{only} on $\lambda$:  
\begin{align}
    J(\lambda) := D\textbf{G}(\textbf{p}^{*}_{\lambda}(\textbf{s})) \label{eq: Jmatrix}
\end{align}
Then $\lambda_{GCV}$ for nonlinear inverse problems is defined as the minimizer of a function involving $J(\lambda)$ \citep{haber2000gcv}:

\begin{align}
    &\lambda_{GCV}(\textbf{s}) = \underset{\lambda \geq 0}{\textsf{argmin}}\:\textsf{GCV}(\lambda)  \\
    &:=\underset{\lambda \geq 0}{\textsf{argmin}}\:\frac{\|G\left(\textbf{p}^{*}_{\lambda}\left(\textbf{s}\right)\right) - \textbf{s}_i\|_{2}^{2}}{\textsf{Tr}\left[\left(I - J(\lambda)(J(\lambda)^{\top}J(\lambda) + \lambda I)^{-1}J(\lambda)^{\top}\right)^2\right]} 
    \label{eq: GCV function}
\end{align}

The concatenated ILR input vectors are produced according to Eq. \eqref{eq: concatenated vector} with $\lambda_{GCV}(\textbf{s})$ used place of $\lambda(\textbf{s})$. The resulting NN trained to minimize the loss Eq. \eqref{eq: NDReg NN loss function} is termed $\textsc{(ND, Reg)}_{\textsc{GCV}}$ (Figure \ref{fig: overall workflow}). Note that constructing training data for either the NN or GCV-based approach requires grid-based optimization for obtaining $\lambda_{oracle}(\textbf{s}_i)$. However, once training is completed, $\lambda_{NN}$ requires no further grid-based searches for operation, while 
obtaining $\lambda_{GCV}(\textbf{s}_i)$ still requires this at the testing, or implementation, step, in order to solve Eq. \eqref{eq: GCV function}.

\section{Results}
\label{sec:results}

\subsection{Parameter Estimation}
\label{subsect: param estimation}

\subsubsection{Estimating $\lambda_{oracle}$}

Learning the signal-dependent optimal Tikhonov regularization parameter $\lambda_{oracle}$ is a problem of independent interest in inverse problems. Here we present our results for estimation of the optimal $\lambda$ for the three-parameter problem Eq. Eq. \eqref{eq:3}. 

\begin{figure}[h]
    \centering
    \includegraphics[width=\textwidth]{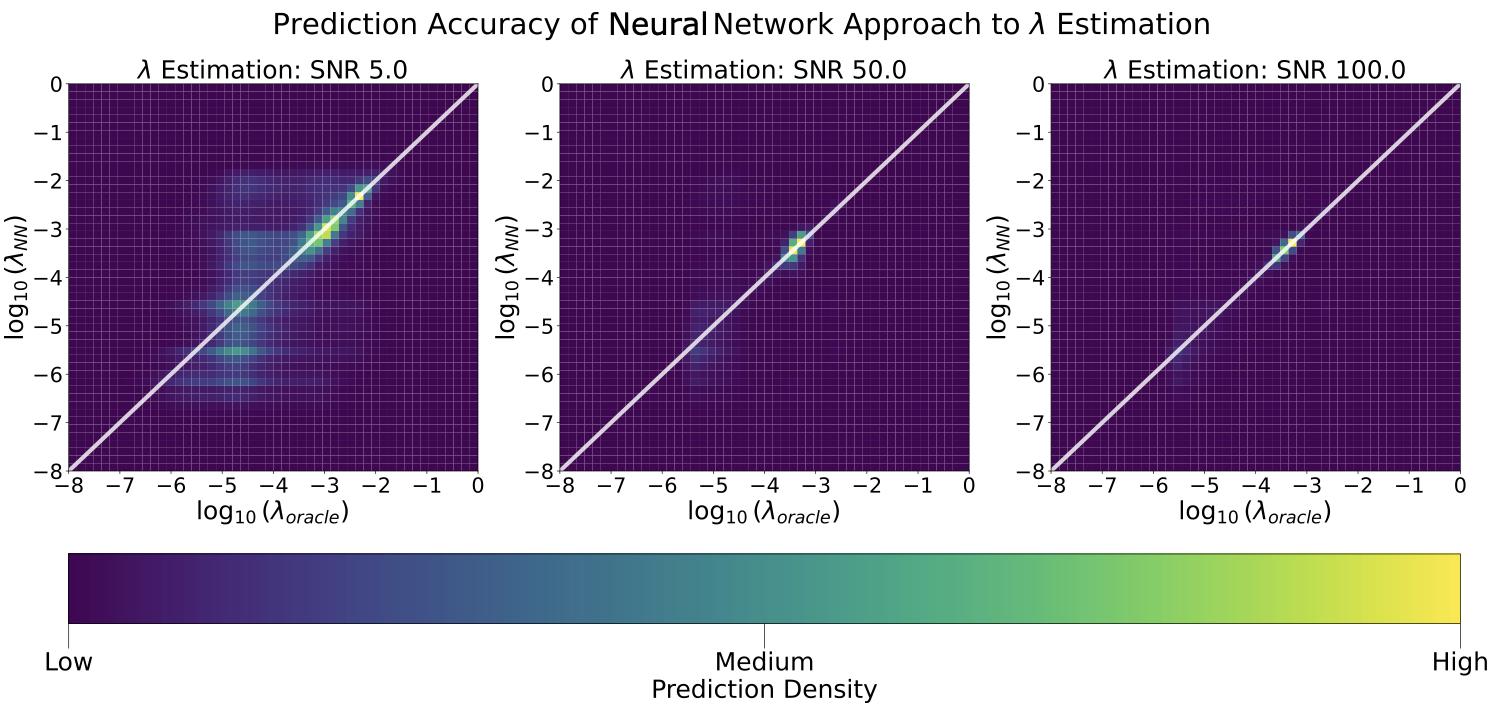}
    \caption{3D Histogram showing prediction density of the $\lambda$-selection neural network versus the $\lambda_{\text{oracle}}$ approach detailed in Algorithm \eqref{alg:datasetGen}. The white line represents the $y=x$ line, indicating the benchmark for perfect prediction. Left to right: SNRs 5 (low), 50 (medium), 100 (high).}
    \label{fig:3D_Lam}
\end{figure}

Figure \ref{fig:3D_Lam} shows that the $\lambda$-selection network $\lambda_{NN}$ when trained on the $L^1$ loss for solving the bilevel problem Eq. \eqref{eq: upper level} closely approximates $\lambda_{oracle}$ over unseen examples of noisy signals $\textbf{s}$. As expected, estimation quality improves with values clustering more closely to the diagonal as SNR increases. 

 In Table \ref{tab:w_1 estimation results}, we compare $\lambda_{NN}$ and $\lambda_{GCV}$ in terms of reproducing the distribution of the values of $\lambda_{oracle}$ as quantified by the \emph{earth mover's distance}. We find that the distribution of the ensemble $\{\lambda_{NN}(\textbf{s})\}$ is closer to the true ensemble than the distribution of $\{\lambda_{GCV}(\textbf{s})\}$ (Table \ref{tab:w_1 estimation results}).

\begin{figure}[h]
    \centering
    \includegraphics[width=1.0\textwidth]{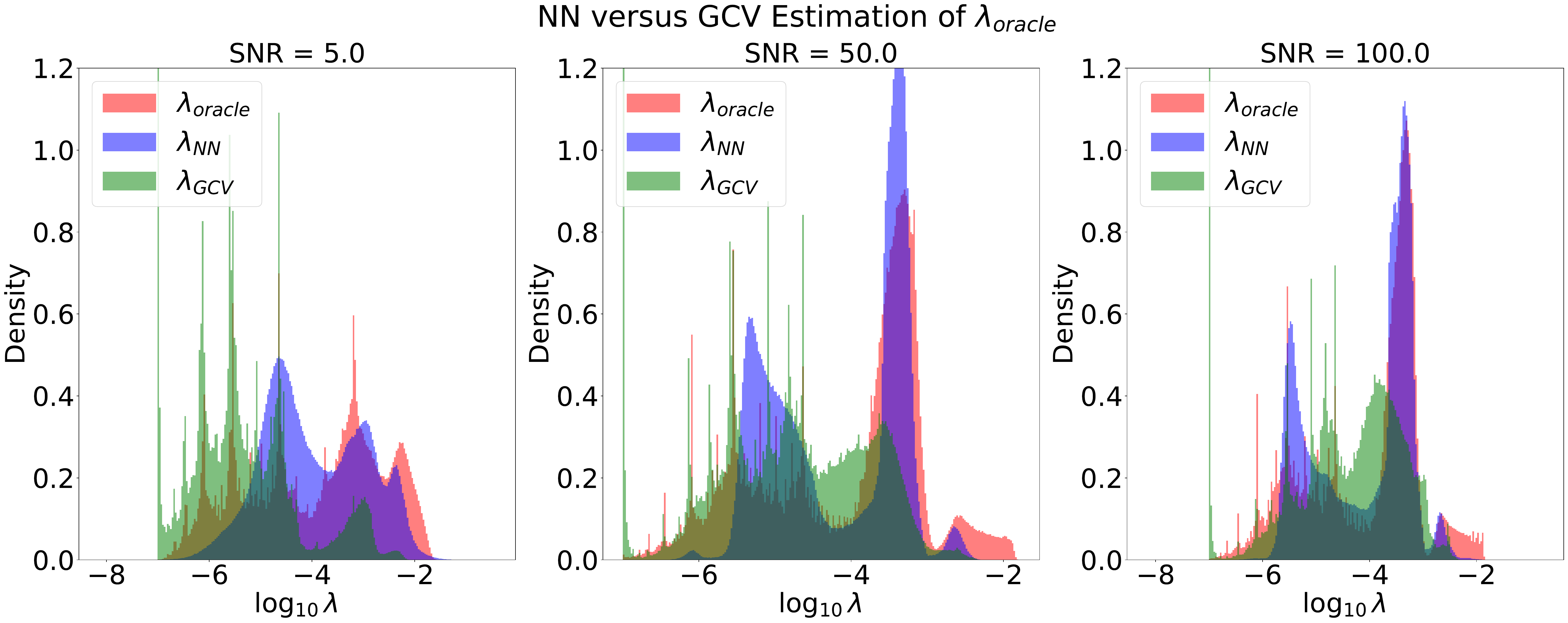}
    \caption{Left to right: The true $\lambda$ distribution (orange) overlaid with the distributions of $\lambda_{NN}$ and $\lambda_{GCV}$. The true distributions at the low SNR level are more multimodal and have greater spread, with GCV providing superior estimates as compared to NN $\lambda$ selection. This is particularly seen at lower values of $\lambda$, where little to no regularization is needed. At the higher SNR's, the optimal $\lambda$ values are concentrated around a small number of modes, with approximation of $\lambda_{oracle}$ by the NN now exhibiting performance superior to GCV.}
    \label{fig:NNvsGCV_distribution}
\end{figure}

In Figure \ref{fig:NNvsGCV_distribution} we disaggregate the values in Table \ref{tab:w_1 estimation results} by comparing the ensembles of $\{\lambda_{NN}(s)\}, \{\lambda_{GCV}(s)\},$ and $\{\lambda_{oracle}(s)\}$ values over the testing set. 
    Despite $\lambda_{NN}$ outperforming $\lambda_{GCV}$ according to the $W_1$ distance overall, $\lambda_{GCV}$ replicates the distribution of $\lambda_{oracle}$ more faithfully in the regime of low SNR and small $\lambda$, $\lambda \in [10^{-7}, 10^{-4}]$. These results can be described by histogram multimodality and spread in this regime, where the $\lambda_{oracle}$ distribution has greater spread and exhibits multimodality that is more accurately captured by GCV than by the NN $\lambda$ estimation, which tends to provide a single dominating mode for these lower values of $\lambda$ (left panel, Figure \ref{fig:NNvsGCV_distribution}). Thus, although the median $\lambda_{oracle}$ is larger for low SNR as expected, the strategy for estimating $\lambda_{oracle}$ accurately over the ensemble of noisy signals $\textbf{s}$ depends on higher order statistics of the $\lambda_{oracle}$ distribution. This suggests that in real applications, an initial noise-estimation step may be of benefit when deciding between use of $\lambda_{GCV}$ and $\lambda_{NN}$. 
    \begin{table}[h]
    \centering
    \begin{tabular}{c|c|c|c}
        \hline \\[0.06em]
        \textbf{SNR} & 5.0 & 50.0 & 100.0 \\
        \hline \\[0.06em]
        $W_{1}(\lambda_{NN}(\textbf{s}), \lambda_{oracle}(\textbf{s}))$ & 0.348 & 0.246 & 0.145 \\
        \hline\\[0.06em]
        $W_{1}(\lambda_{GCV}(\textbf{s}), \lambda_{oracle}(\textbf{s}))$ & 1.386 & 0.862 & 0.560 \\
        \hline
        \hline
    \end{tabular}
    \caption{The Wasserstein-1, or earth mover's distance (EMD), between the ensembles of $\lambda$ values produced by the NN (upper row) and GCV (lower row). Note that the NN outperforms GCV in matching the distribution to $\lambda_oracle$ for every SNR level. However, the multiple modes towards low values of $\lambda$ are captured more accurately by GCV.}
    \label{tab:w_1 estimation results}
\end{table}

\subsubsection{Accuracy of ILR and Alternative Methods for $c_1$ Estimation}

We have defined the two ILR NN, $\textsc{(ND, Reg)}_{\textsc{GCV}}$ and $\textsc{(ND, Reg)}_{\textsc{NN}},$ above. For comparison with conventional NN estimation, we duplicate the native noisy input vector to establish an input of the same length as the ILR NN's and thereby eliminate input signal length as a potential confounder. We call this NN $\textsc{(ND, ND)}$. Additional standards of comparison are the conventional NLLS estimation method, as well as the Tiknonov-regularized (TR-NLLS) approach in which regularization is applied to this parameter estimation problem; in our case, we have implemented this with selection of $\lambda$ via GCV. 

Results for estimation of $c_1$, representing MWF in our motivating application, are shown in Table \ref{tab:c1_parameter_estimation_results}. Several points of interest are evident. First,  $\textsc{(ND, Reg)}_{\textsc{GCV}}$ and $\textsc{(ND, Reg)}_{\textsc{NN}}$ perform essentially indistinguishably. 
However, both of these ILR-based NN outperform the NN without ILR, $\textsc{(ND, ND)}$, across all SNR levels. This is our main result. We also note (not shown) that the NN architecture without ILR and without duplication of the input vector, deemed $\textsc{(ND)}$, performs virtually identically with $\textsc{(ND, ND)}$. Further, the most conventional analysis, NLLS, exhibits performance substantially inferior to any of the NN's. In addition, the non-conventional method of TR-NLLS, in spite of the introduction of bias through regularization, provides sufficient stabilization that its overall performance is markedly superior to that of NLLS. In appendix B, we provide a more detailed study of $\textsc{(ND, Reg)}_{\textsc{NN}}$ and $\textsc{(ND, Reg)}_{\textsc{GCV}}$ by disaggregating the RMSE over all parameter combinations appearing in the testing data. 
\begin{table}[h]
    \centering
    \begin{tabular}{@{\extracolsep\fill}lccc@{\extracolsep\fill}}
        \toprule
        \textbf{SNR} & \textbf{5.0} & \textbf{50.0} & \textbf{100.0} \\
        \midrule 
        
        \textbf{(ND, Reg)$_{NN}$ $\text{RMSE}_{c_1}$} & 0.1652 & 0.0931$^*$ & 0.0796$^*$ \\
        \textbf{(ND, Reg)$_{GCV}$ $\text{RMSE}_{c_1}$} & 0.1643$^*$ & 0.0945 & 0.0817 \\
        \textbf{(ND, ND) $\text{RMSE}_{c_1}$} & 0.1671 & 0.1079 & 0.0935 \\
        \textbf{TR-NLLS $\text{RMSE}_{c_1}$}, $\lambda_{oracle}$ & 0.2328 & 0.2092 & 0.2078 \\
        \textbf{TR-NLLS $\text{RMSE}_{c_1}$}, $\lambda_{GCV}$ & 0.3687 & 0.3487 & 0.3347 \\
        \textbf{NLLS $\text{RMSE}_{c_1}$} & 0.3844 & 0.3471 & 0.3453 \\
        \bottomrule 
    \end{tabular}
    \caption{Root mean squared error (RMSE) in $c_1$ for each indicated parameter estimation method. * indicates superior performance for the given SNR category.}
    \label{tab:c1_parameter_estimation_results}
\end{table}

\subsection{Myelin water fraction estimation}
We now apply ILR to the estimation of the MWF from magnetic resonance relaxometry (MRR) signals from the human brain. A multi-spin-echo sequence with 64 values of TE was applied as outlined in the Appendix, with the sampling scheme leading to the decay signal as described in Eq. \eqref{eq:1}. 

However, we note that while the white matter of the brain, rich in myelin content, is expected to exhibit biexponential decay, the grey matter, with little-to-no myelin, would be better described by a monoexponential decay. For the latter,  Eq. \eqref{eq:1} represents an underdetermined model, and no amount of regularization or other manipulation will permit meaningful parameter values to be extracted. The determination of two exponentials will be enforced by the model, with their fractions $c_1$ and $c_2$ noise-dependent and therefore random. Therefore, we wish to apply our analysis only to truly biexponential signals. Accordingly, we evaluate each pixel using the Akaike information criterion (AIC) \citep{james2013introduction}, and analyze signals only from pixels that are better described as biexponential rather than monoexponential. We note that the distribution of the AIC-defined mono- and bi-exponential pixels closely follows the general pattern of cortical grey and interior white matter in the brain.  

\begin{figure}[ht]
    \centering
    \includegraphics[width=0.7\textwidth]{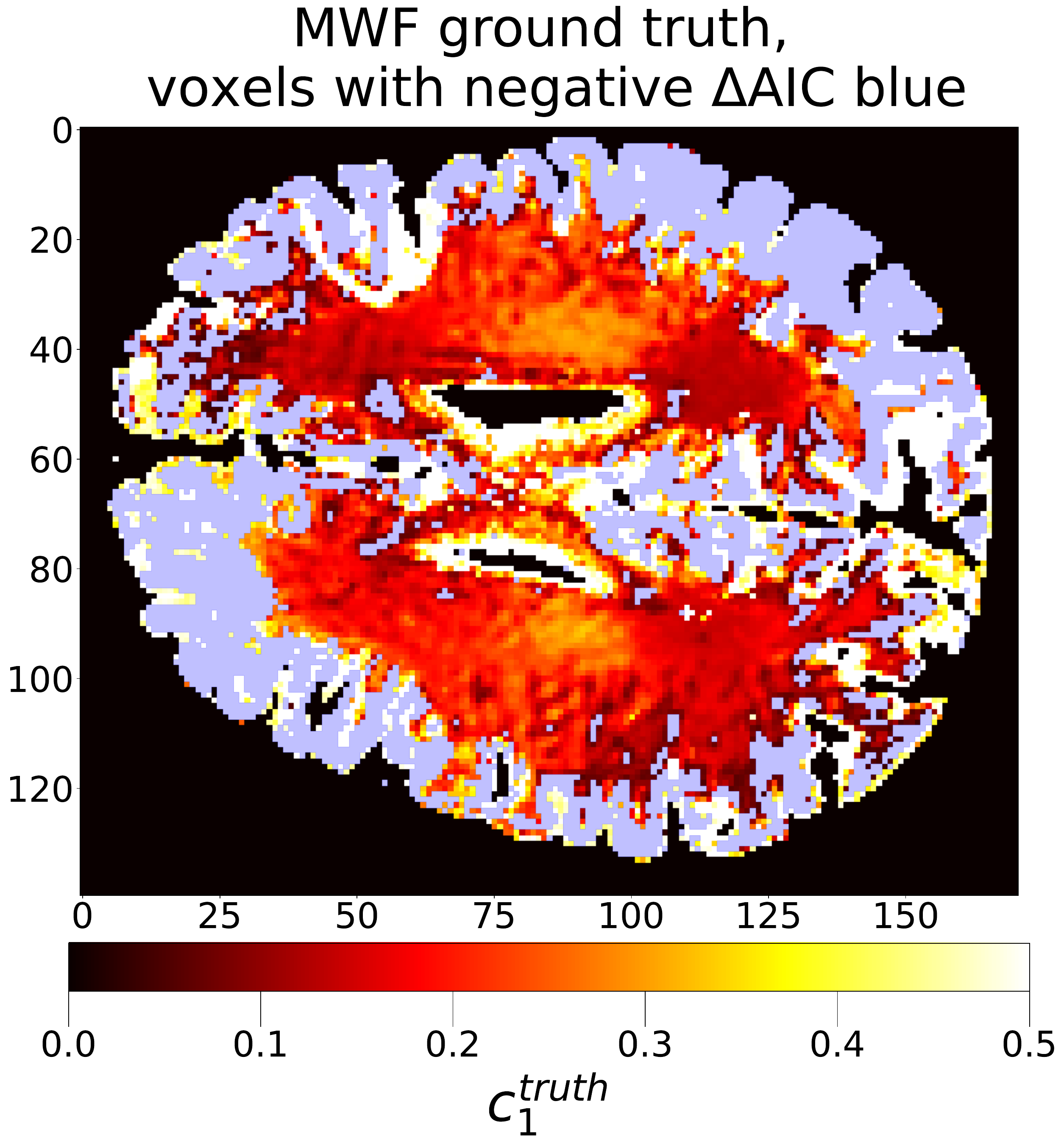}
    \caption{Comparison standard for $c_1$ for pixels determined to be biexponential according to the AIC (color bar); purple pixels are those determined to be of a monoexponential character. Each voxel in the 288 $\times$ 288 image of the brain is characterized by a noisy signal $\textbf{s}$ with 64 acquisition times. Each of these voxel signals was evaluated according to a 5-parameter model after NESMA filtering\citep{bouhrara2018use}. Estimates for $c_1$ were taken as a comparison standard.}
    \label{fig:MWFtrue}
\end{figure}

With this, the time-series for each voxel under consideration is a single noisy biexponential curve with parameters to be estimated via ILR with signal-dependent $\lambda$ selection. We estimated ground truth parameters through the application of the nonlocal estimation of multispectral magnitudes (NESMA) denoising filter \citep{bouhrara2018use}, followed by NLLS. We include additional information regarding the acquisition of brain data in Appendix E. The estimates for $c_1$ are visualized in Figure \ref{fig:MWFtrue}.

As suggested in simulations, high quality performance of the (ND, Reg) networks is predicated on proper selection of the regularization parameter $\lambda$ using either GCV or $NN$; in our case, this means selection of $\lambda$ that is similar to $\lambda_{oracle}$, or at least that provides comparable performance. Furthermore, we illustrated that the choice of GCV or $NN$ can be made based on estimation of SNR. Using a comparable method of SNR quantification, we found median SNR of $\approx 25.0$ on the biexponential pixels identified by the AIC, suggesting that the underlying distribution of $\lambda_{oracle}$ was likely to be spread over orders of magnitudes with a large number of modes, as in Figure \ref{fig:NNvsGCV_distribution}. Table \ref{tab:c1_parameter_estimation_results} indicates that superior results may be obtained from $\textsc{(ND, Reg)}_{\textsc{GCV}}$ as compared to $\textsc{(ND, Reg)}_{\textsc{NN}}$.

\begin{figure}[h]
        \centering
        \includegraphics[width=\linewidth]{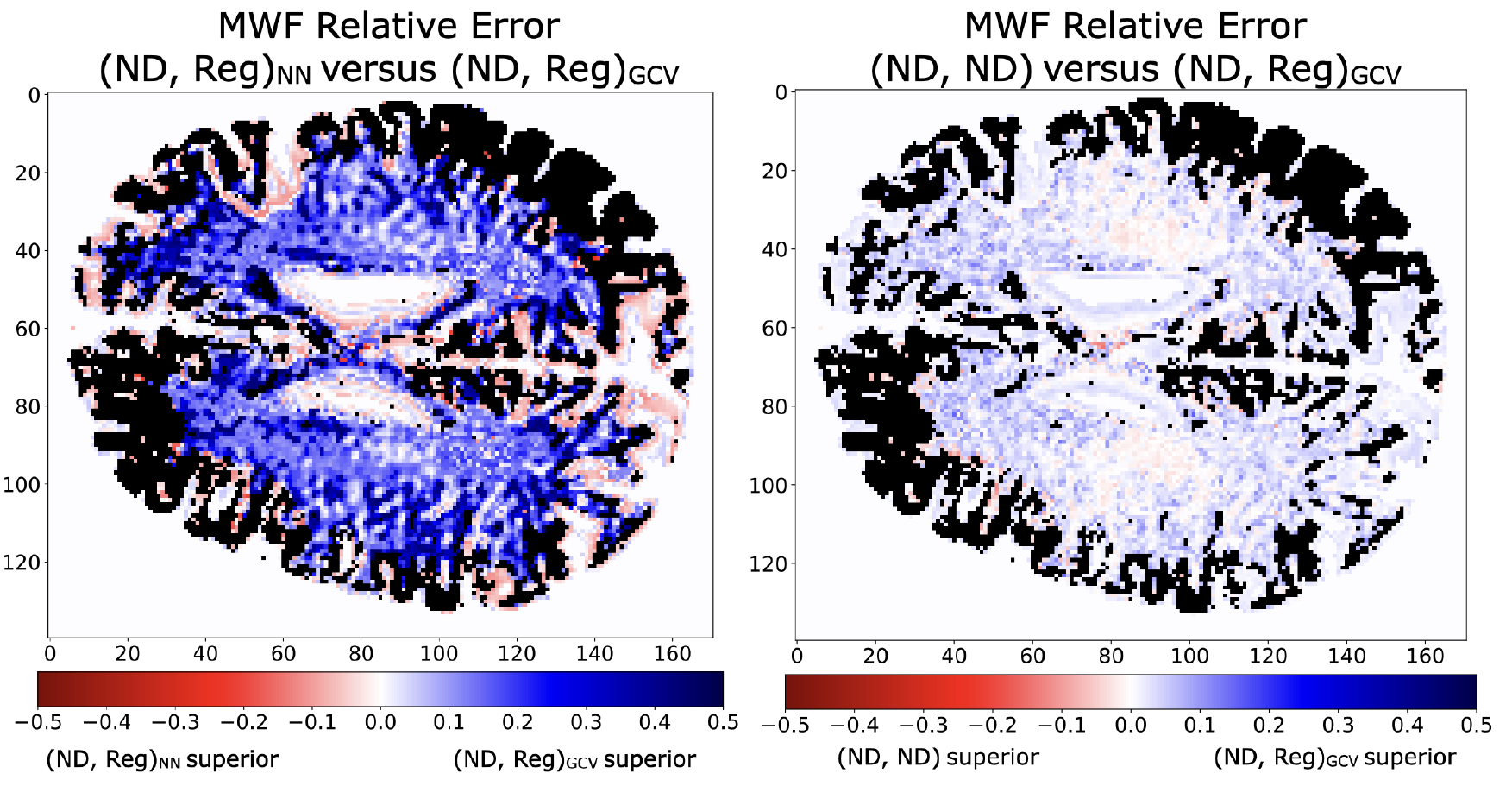}
        \caption{Left: Comparison of $c_1$ estimation according to the two methods of $\lambda$ selection. The metric illustrated is the difference in RMSE values $RMSE(\textsc{(ND, Reg)}_{\textsc{NN}}) - RMSE(\textsc{(ND, Reg)}_{GCV})$ calculated over pixels with a biexponential character. As seen, $\textsc{(ND, Reg)}_{GCV}$ is superior overall.  Right: Comparison of the better of the two $\lambda$ selection methods, $GCG$, with the NN without incorporating ILR, based on $RMSE(\textsc{(ND, ND)}) - RMSE(\textsc{(ND, Reg)}_{\textsc{NN}})$. We find that $\textsc{(ND, Reg)}_{\textsc{GCV}}$ exhibits superior performance overall.}
        \label{fig:brain_ILR}
    \end{figure}

Indeed, the left panel of Figure \ref{fig:brain_ILR} shows that $\textsc{(ND, Reg)}_{\textsc{GCV}}$ is in general superior to  $\textsc{(ND, Reg)}_{\textsc{NN}}$. This is also consistent with Figure \ref{fig:NNvsGCV_distribution}, suggesting that $\lambda_{GCV}$ may more accurately reproduce the modes of the distribution of $\lambda_{oracle}$; this capability is not captured by either the $L^1$ loss or the earth mover's distance. Therefore, $\lambda_{GCV}$ may be preferable for use with ILR with limited SNR.  

Adopting $\textsc{(ND, Reg)}_{\textsc{GCV}}$ as the method of choice for this application, we compared its performance with that of the $\textsc{(ND, ND)}$ network, as shown in the right panel of Figure \ref{fig:brain_ILR}. We find that $\textsc{(ND, Reg)}_{\textsc{GCV}}$ outperforms $\textsc{(ND, ND)}$ nearly \emph{everywhere}, indicating the efficacy of ILR to enhance NN-based MWF estimation. It is notable that optimal performance requires combining this NN with a classical technique for selection of the regularization parameter $\lambda$, $\lambda_{GCV}$, rather than selection according to the $\lambda_{NN},$ for generating the augmented part of the noisy input signal. 

\section{Discussion}
\label{sec:discussion}
Combining conventional analytic methodology and domain knowledge with machine learning-based methods has become a common paradigm \cite{zheng2021leveraging, guo2022using, cuomo2022scientific}. Examples include incorporating physical constraints into loss functions \cite{cai2021physics}, seeding classical solvers with machine-learned initializations \cite{zhou2023neural}, and designing individual layers of neural networks around the physics of the problem \cite{villar2021scalars}. Here we have presented a new approach that is an additional example of such a methodology by integrating the classical technique of Tikhonov regularization \cite{burger2004convergence, lorenz2013necessary, flemming2010theory, scherzer2009variational} and bilevel optimization \cite{colson2007overview, dempe2020bilevel} with neural network data augmentation and automated hyperparameter tuning. This leads to a refined approximation of the map from a noisy signal $\textbf{s}$ to a parameter estimate $\widehat{\textbf{p}}$. In particular, we have presented a cascaded approach to parameter estimation with two levels: (1) estimating the hyperparameter $\lambda$ from the noisy signal $\textbf{s}$ followed by (2) estimating the parameters of a decaying exponential using ILR, with the regularized component of the concatenated signal generated using $\lambda$ from the previous step. 

For the first level of analysis, we explored two strategies for recovering $\lambda$ from the noisy signal $\textbf{s}$, which can itself be viewed as a parameter estimation problem. The first strategy involves selection of lambda using GCV \cite{golub1979generalized}, a common, statistically validated, method that does not require an estimate of signal noise level. This approach requires no training, but rather evaluates each signal individually and estimates the optimal lambda by what is in effect a leave-one-out cross-validation. The second strategy is a neural network \cite{Afkham} trained with a loss function involving $\lambda_{oracle}$.
Both methods involve computational bottlenecks related to the solutions of optimization problems.  The GCV estimate requires minimization over $[0, \infty)$ for the GCV functional, which can be difficult to optimize with gradient methods, while the NN approach requires the generation of $\lambda_{oracle}$ values through an expensive grid search strategy sweeping over several orders of magnitude. Nonetheless, we see that a four-layer CNN can approximate the map $\textbf{s} \to \lambda(\textbf{s})$ accurately, suggesting that even a modest number of layers can suffice for approximating solutions to this highly unstable problem. We have also demonstrated that the $L^1$ loss is an effective tool to approximate not only the map $\textbf{s} \to \lambda_{oracle}(\textbf{s})$ but also the \emph{ensemble} of $\lambda_{oracle}$ considered as a probability distribution. The $\lambda_{NN}$ outperforms $\lambda_{GCV}$ with respect to both of these error metrics. However, we also find that at lower SNR, $\lambda_{GCV}$ appears to be superior in deriving parameters from noisy signals that do not require substantial regularization. The effect of NN depth, loss functions, and selection of activation functions on estimation of optimal $\lambda$ values remains an open question.  

For the second level of our analysis pipeline, we used $\lambda_{GCV}$ and $\lambda_{NN}$ to construct a regularized signal which was concatenated with the noisy signal, with this concatenated construct used to train the ILR networks $\textsc{(ND, Reg)}_{\textsc{NN}}$ and $\textsc{(ND, Reg)}_{\textsc{GCV}}$. Both of these $\textsc{(ND, Reg)}$ networks, differing only in the manner in which $\lambda$ was selected, attained over 5--10\% or greater improvements in accuracy compared to the basic ND NN and TR-NLLS estimation. This indicates that ILR is a promising approach for overcoming the limitations of traditional methods when performing parameter estimation from noisy decay signals. In particular, our results suggest that ILR may significantly improve the accuracy and reliability of parameter estimation problems in magnetic resonance relaxometry, a topic of great current interest for gaining insights into the pathophysiology of the central nervous system \cite{faizy2018age,kolind2012myelin,mackay2016magnetic}.

We also explored the effect of noise on the accuracy of parameter estimates produced by the ILR networks with NN $\lambda$ selection. As expected, the accuracy of parameter recovery is sensitive to SNR, further underscoring the fundamental importance of data quality in MRR. In our setting, noise affects two stages of the parameter estimation; one of these is the more conventional effect of noise on the parameter estimation NN, while the other is the error in estimation of the regularization parameter $\lambda$ given by the $\lambda$-NN.  This is reflected in the inaccuracy of the $\lambda$-NNs in estimating the oracle distribution of $\lambda$. In fact, we find that as the SNR increases, the multimodality of this distribution decreases (Figure \ref{fig:NNvsGCV_distribution}), leading to enhanced NN estimation. This observation provides insight into the mechanism through which noise affects the distribution of recovered $\lambda$: noise appears to promote clustering in the distribution of $\lambda$ values. Understanding relationships such as this between the distribution of the data and Tikhonov regularization parameters has been a long-standing topic in inverse problems and statistical learning theory \cite{evgeniou2002regularization,cherkassky2009another}. Theoretical results have been developed for cases in which the forward map in the inverse problem is linear \cite{alberti2021learning,ehrhardt2023optimal} and well-conditioned \cite{neubauer1988posteriori,engl1987optimal}, or when the underlying noise distribution is well-behaved \cite{golub1979generalized,scherzer1993optimal}. Here we significantly relax these constraints and provide empirical results for the case in which the forward map is nonlinear and the noise is Rician. 

The optimal construction of nonlinear models to capture the complexities of tissue remains an open question. In particular, we find in Figure \ref{fig:brain_ILR} that although input layer regularization improves upon the pure deep learning NN approach, the construction of the regularized curves in the low SNR regime is better performed via a classical method rather than through a NN. This method then represents a hybrid between classical and deep learning approaches for parameter estimation, where the final MWF is estimated using ILR, but where the augmented component of the signal input to the ILR network may be generated using classical methods, namely, TR-NLLS with GCV-selected $\lambda$.

The present analysis does not incorporate the spatial correlations between pixels which exist in any natural image. Instead, our $\lambda_{NN}$ networks are trained to process data arising from individual decay curves. Thus, although $\lambda_{NN}$ provides meaningful estimates of $\lambda_{oracle}$, further optimization of its architecture, including potential modifications of the penalty function to penalize rapid spatial variation, may improve results. The challenges in this approach are to avoid the introduction of bias and of blurring. 

In actual brain data, there is no gold standard for parameter estimation, and surrogates for actual underlying values must be used instead. In the present case, we applied an effective noise-reduction filter to obtain the benchmark results shown in Figure \ref{fig:MWFtrue}. Nevertheless, application to an actual in vivo system is an important part of demonstrating the potential utility of the ILR approach, especially since actual data is unlikely to strictly obey any imposed signal model. 

The data-augmentation approach we have developed and implemented demonstrates a substantial improvement in the accuracy of parameter estimation for the biexponential decay signal. Nevertheless, there are certain limitations to our study which may form the basis for additional investigation.  Although we have carefully tuned the hyperparameters for our various NN's, this remains something of a black art in the literature and results could change somewhat with additional tuning. We have elected to perform our investigations using the Rician noise model as the basic model in MRI, and there is nothing in our analysis that would be specific to this choice. Nevertheless, investigation of other models, and in particular the Gaussian, would be of interest. In addition, as noted above, we have demonstrated our method on actual brain data, but these results are somewhat inferential, since there is no true gold standard for underlying parameter values. We also note that one of our comparison methods, TR-NLLS, is highly nonstandard. We have however found this to be an effective modification of NLLS in many settings and are exploring this more systematically. Further, we applied the AIC to define pixels for which signal regularization makes sense; this appears to be a novel approach in the literature and has proved to be very effective here. However, there are several choices, including the F-test, the Bayesian information criterion, and reduced chi-squared, to address this issue, and another choice may exhibit superior performance. Lastly, we highlight the fact that our work as presented here explores only the biexponential model, with emphasis on determination of the $c_1$ parameter. Other signal models and parameter estimation problems should also be evaluated in this framework. 

In summary, we have addressed the recovery of the parameter $c_1$ in the biexponential model in Eq. \eqref{eq:Biexp} for MWF estimation in the central nervous system, a topic of ongoing research interest.  On both simulated data and actual brain data, we find ILR to offer a significant improvement in parameter estimation accuracy.

\section*{Acknowledgments}
This work was supported in part by the Intramural Research Program of the National Institute on Aging of the NIH





\subsection*{Conflict of interest}

The authors declare no potential conflict of interests.

\section*{Data availability}
The code for generating the data and figures has been made available at \url{https://github.com/ShashankSule/ILR-for-MWF}.

\appendix
\section{Dataset generation}
\label{subsec: datagen}
We generate the training data used for $\lambda_{NN}$, $\textsc{(ND, Reg)}_{\textsc{NN}}$, and $\textsc{(ND, Reg)}_{\textsc{GCV}}$ as follows. For each parameter vector $\textbf{p}$ in a uniformly spaced grid of possible parameter values specified in Table \ref{tab2}we generate a noiseless signal $\textbf{G}(\textbf{p})$ using (\ref{eq:Biexp}). Every noiseless signal is then corrupted with 1000 independent and identically distributed (iid) copies of Rician noise to generate noisy signals resembling experimental MRI data. The noisy signals are normalized to have the initial signal amplitude be 1.0 at the minimum echo time $TE_0 = 8.0$. Finally, we assemble all the noisy signals over all the parameter vectors into the set of training inputs denoted by $\{\textbf{s}_i\}$. Then the parameters used for generating $\textbf{s}_i$ will be indexed accordingly and assembled into the set of training outputs given by $\{\textbf{p}_i\}$. To obtain $\lambda_{oracle}(\textbf{s}_i)$, we perform a grid search over $\lambda \in [10^{-7}, 10^{3}]$ for finding minimizers to \eqref{eq: upper level}. For obtaining $\lambda_{GCV}$, we employ another grid search in the range $\lambda \in [10^{-7}, 10^{3}]$ to find the minimizers in \eqref{eq: GCV function}. Our procedure is summarized in Algorithm \eqref{alg:datasetGen}.
 
\begin{table*}[h]
\centering
\caption{Parameters defined for dataset creation}.\label{tab2}%

\begin{tabular*}{\textwidth}{@{\extracolsep\fill}p{0.4\textwidth}p{0.1\textwidth}p{0.4\textwidth}@{\extracolsep\fill}}
\toprule
\textbf{Parameter} & \textbf{Symbol}  & \textbf{Value}\\
\midrule
Number of signal acquisition times & $\textit{N}_t$ & 32 \\ 
Lower and upper bounds on regularization parameter search space & $\log_{10}(\lambda_{oracle})$ & \: \: -7,3 \\ 
Initial guess for $\lambda_{oracle}$-solver (regularized-nonlinear least squares solver) & \textbf{z} & $(T_{2,1}^{true}, T_{2,2}^{true}, c_{1}^{true})$ \\
Lower and upper bounds on target fraction of more rapidly decaying component & $c_1^l,c_1^h$ & \begin{minipage}[t]{0.5\textwidth}
    [0.0,0.6] \: (Training),  \newline [0.0, 0.5] \: (Validation \& Testing)
\end{minipage} \\
Fraction of more slowly decaying component & $c_2$ &$1.0-c_1$ \\
Lower and upper bounds on target $T_{2,1}$ & $T_{2,1}^l, T_{2,1}^h$ & \begin{minipage}[t]{0.5\textwidth}
    [1.0, 50.0] \: (Training), \newline [5.0, 45.0] \: (Validation \& Testing)
\end{minipage}\\
Lower and upper bounds on target $T_{2,2}$ & $T_{2,2}^l, T_{2,2}^h$ & \begin{minipage}[t]{0.5\textwidth}[40.0, 225.0] (Training), \newline [45.0, 200.0] (Validation \& Testing)\end{minipage}\\
Number of uniformly spaced $T_{2}$ and $c_1$ values & $\textit{N}_p$ & 20 (per parameter) \\
Range of signal acquisition times & & $[8.0, 256.0]$\\
Number of noise realizations & $N_{\omega}$ & 1000 (Training), 200 (Validation), 400 (Testing)\\
\bottomrule
\end{tabular*}

\end{table*}

\begin{algorithm}
\centering
\caption{Dataset Generation}\label{alg:datasetGen}
\begin{algorithmic}[1]
\State Set dataPurpose \Comment{dataPurpose is training, validation, or testing}
\State Initialize parameters from Table \ref{tab2}. 
\State \hspace{\algorithmicindent}dataPurpose $T := \{(c_1, T_{2,1}, T_{2,2}) \in [c_1^l,c_1^h] \times [T_{2,1}^l, T_{2,1}^h] \times [T_{2,2}^l, T_{2,2}^h] : T_{2,1} \leq T_{2,2} \}$

\State \hspace{\algorithmicindent}SNR
\State Initialize empty list for samples: samples = List$(N_T \cdot N_\omega)$
\State Seed random number generator: seed(dataPurpose)
\For{$[c_1, T_{2,1}, T_{2,2}] \in T$}
    \State Compute noiseless signal: $\textbf{G}(c_1, T_{2,1}, T_{2,2}) := (c_1\mathrm{e}^{\frac{-t_n}{T_{2,1}}} + (1.0-c_1)\mathrm{e}^{\frac{-t_n}{T_{2,2}}})^N_{n=1}$
    \For{noise realization $k=1,\ldots,N_\omega,$}
        \State Get noise realization $\eta,\xi \sim \mathcal{N}(0, \frac{1}{SNR^2}I_N)$ 
        \State Add Rician noise to noiseless signal: 
        \newline
        \begin{equation}
            \textbf{s} = \sqrt{(G(\textbf{p}) + \xi)^2 + \eta^2}
        \end{equation}
        
        \For{regularization parameter $\lambda \in [10^{-7},10^{3}]$}
            \State Solve Tikhonov regularized NLLS problem with \texttt{curve\_fit}:
                \State \hspace{\algorithmicindent}$\textbf{p}_\lambda^* \equiv \textsf{argmin}_{\textbf{p} \in \mathbb{R}_+^2}({\lVert{\textbf{G(p)} - \textbf{s}}\rVert}^2_2 + \lambda^2\lVert{\textbf{p}}\rVert_2^2).$
        \EndFor
        \State $\lambda_{oracle} := \textsf{argmin}_{\lambda}((\textbf{p}_\lambda^* - \textbf{p}_{opt})^2)$
        \State Append $(G, \lambda_{oracle})$ to samples 
    \EndFor
\EndFor
\end{algorithmic}
\end{algorithm}

\section{NN training}

In this paper we trained four distinct NNs:  
\begin{enumerate}
    \item $\lambda_{NN}$ for estimating $\lambda_{oracle}$ from the noisy signal $\textbf{s}$,
    \item $\textsc{(ND, Reg)}_{\textsc{NN}}$ for estimating $\textbf{p}_{true}$ from $(\textbf{s}, G(\textbf{p}_{\lambda_{NN}(\textbf{s})}(\textbf{s}))$,
    \item $\textsc{(ND, Reg)}_{\textsc{GCV}}$ for estimating $\textbf{p}_{true}$ from $(\textbf{s}, G(\textbf{p}_{\lambda_{GCV}(\textbf{s})}(]\textbf{s}))$, and
    \item $\textsc{(ND, ND)}$ for estimating $\textbf{p}_{true}$ from $(\textbf{s}, \textbf{s})$.
\end{enumerate}

\subsection{Training $\lambda_{NN}$}
Here we outline the procedure to train our $\lambda_{NN}$ to learn the optimal regularization parameter. The process is detailed in Algorithm \ref{alg:Lambda-Selection NN}. The general flow is (1) Initialize the NN model, (2) Train the network via supervised learning, (3) Validate the model's performance.   
\begin{algorithm}[h]
\caption{$\lambda$-selection Neural Network}\label{alg:Lambda-Selection NN}
\begin{algorithmic}[1]
\State Load training set $\mathbb{TR}$ and validation set $\mathbb{V}$
\State Initialize neural network (NN) hyperparameters:
\State \hspace{\algorithmicindent}number of epochs $N_e=30$
\State \hspace{\algorithmicindent}samples per batch $B=64$
\State \hspace{\algorithmicindent}learning rate $\ell=0.0001$
\State Initialize neural network: $\mathbf{F}^{(0)} := \mathbf{F}(\cdot, \boldsymbol{\theta}^{(0)})$
\State Initialize minimum validation metric $v := \inf$ \Comment{So large it is immediately overwritten}
\For{epoch $e = 1, 2, \ldots, N_e$}
    \For{batch $b = 1,2,\ldots, N_b$} \Comment{$N_b = N_{\mathbb{TR}} / B$}
        \State Randomly acquire batch $b$ of samples from the training set $\mathbb{TR}$: $(\{\mathbf{X_k}\}, \{\mathbf{Y}_k\})_{k=1}^{B}$
        \State Set gradient of loss to zero
        \State Predict regularization parameters: $\{\lambda_{NN}\}_{k=1}^{B} = \{\mathbf{F}^{(e-1)} (\mathbf{X_k}) \}_{k=1}^{B}$
        \State Compute $L_1$ loss: $L(\{\lambda_{NN}\}_{k=1}^{B}, \{\mathbf{Y}_k\})_{k=1}^{B}) := {\frac{1}{B}\sum_{k=1}^{B} \lVert (\lambda_{NN})_k - \mathbf{Y}_k \rVert}_1$
        \State Compute loss gradient w.r.t. NN weights and biases $\boldsymbol{\theta}$ with backpropagation: $\triangledown_{\boldsymbol{\theta}}L$
        \State Update NN weights and biases: $\boldsymbol{\theta}^{(e-1)} \gets \texttt{Adam}(\boldsymbol{\theta}^{(e-1)}, \triangledown_{\boldsymbol{\theta}}L, \ell)$ 
    \EndFor
    
    \State $\boldsymbol{\theta}^{(e)} \gets \boldsymbol{\theta}^{(e-1)}$
    \State Compute validation metric $v^{(e)} :=  \frac{1}{N_\mathbb{V}} \sum_{(\mathbf{X}, \mathbf{Y}) \in \mathbb{V}} {\lVert \mathbf{F}^{(e)}(\mathbf{X}) - \mathbf{Y} \rVert}_1$
    \If{$v^{(e)} < v$}
        \State Update running minimum validation metric: $v \gets v^{(e)}$
        \State Update running best weights and biases: $\boldsymbol{\Theta} \gets \boldsymbol{\theta}^{(e)}$
    \EndIf
\EndFor
\State \Return Write the trained NN weights and biases to disk: save($\boldsymbol{\Theta}$)
\end{algorithmic}
\end{algorithm}

\subsection{Training ILR networks and (ND, ND)}
Here we outline the procedure to train our (ND, ND), (ND, Reg)$_{NN}$, and $(ND, Reg)_{GCV}$ models to learn the parameter estimate from the given input. The process is detailed in Algorithm \ref{alg:parameter-estimation NN}. The general flow is (1) Retrieve the regularization parameter estimates via $\lambda_{NN}$ or GCV, (2) Compute the regularized signal, (3) Train the desired model, and (4) Validate the desired model. Note, the (ND, ND) model does not make use of the regularization parameter. Rather, the model's input is the noisy signal concatenated with itself to preserve input size.

\begin{algorithm}
\caption{Parameter Estimation Neural Network}\label{alg:parameter-estimation NN}
\begin{algorithmic}[1]
\State Load parameter estimation training set $\mathbb{TR}$ and validation set $\mathbb{V}$
\State Load $\lambda_{NN}$ or $\lambda_{GCV}$ estimates from disk: $\textbf{D} \gets$ load($\boldsymbol{\Theta}$)
\For{noisy signals $NDs \in [\mathbb{TR}, \mathbb{V}]$}
    \State Get regularization parameter estimates: $\lambda_{estimates} \gets \boldsymbol{D}(NDs)$
    \For{noisy signal $s$ and $\lambda_{est} \in (NDs,  \lambda_{estimates})$}
        \State Solve regularized-NLLS (\eqref{eq:RNLLS}) to get parameter estimates:  $$\boldsymbol{p}_{\lambda_{est}}^{*} \gets \textsf{argmin}_{\textbf{p} \in \mathbb{R}_+^2}({\lVert{\textbf{G(p)} - s}\rVert}^2_2 + \lambda_{est}^2\lVert{\textbf{p}}\rVert_2^2)$$
        \State Reg $\gets \boldsymbol{G}(\boldsymbol{p}_{\lambda_{est}}^{*})$
        
        \State Update dataset $Da$ with new input: $Da \gets (d, Reg)$
    \EndFor
\EndFor
\State Initialize parameter estimation network hyperparameters:
\State \hspace{\algorithmicindent}number of epochs $N_e=30$
\State \hspace{\algorithmicindent}samples per batch $B=64$
\State \hspace{\algorithmicindent}learning rate $\ell=0.0001$
\State Initialize neural network: $\mathbf{F}^{(0)} := \mathbf{F}(\cdot, \boldsymbol{\theta}^{(0)})$
\State Initialize minimum validation metric $v := \inf$ \Comment{So large it is immediately overwritten}
\For{epoch $e = 1, 2, \ldots, N_e$}
    \For{batch $b = 1,2,\ldots, N_b$} \Comment{$N_b = N_{\mathbb{TR}} / B$}
        \State Randomly acquire batch $b$ of samples from the training set $\mathbb{TR}$: $(\{\mathbf{X_k}\}, \{\mathbf{Y}_k\})_{k=1}^{B}$
        \State Set gradient of loss to zero
        
        \State Predict parameter estimates: $\{p_{est}\}_{k=1}^{B} = \{\mathbf{F}^{(e-1)} (\mathbf{X_k}) \}_{k=1}^{B}$
        
        \State Compute Mean Squared Error loss:
        {\small\begin{align}
            &L(\{p_{est}\}_{k=1}^{B}, \{\mathbf{Y}_k\})_{k=1}^{B}) \\
            &:= \frac{1}{B}\sum_{k=1}^{B} (([c_1^{est}, t_{2,1}^{est}, t_{2,2}^{est}]_k - [c_1^{truth}, t_{2,1}^{truth}, t_{2,2}^{truth}]) \times [100.0, 1.0, 1.0])^2
        \end{align}}
        
        \State Compute loss gradient w.r.t. NN weights and biases $\boldsymbol{\theta}$ with backpropagation: $\triangledown_{\boldsymbol{\theta}}L$
        \State Update NN weights and biases: $\boldsymbol{\theta}^{(e-1)} \gets \texttt{Adam}(\boldsymbol{\theta}^{(e-1)}, \triangledown_{\boldsymbol{\theta}}L, \ell)$ 
    \EndFor
    
    \State $\boldsymbol{\theta}^{(e)} \gets \boldsymbol{\theta}^{(e-1)}$
    \State Compute validation metric $v^{(e)} :=  \frac{1}{N_\mathbb{V}} \sum_{(\mathbf{X}, \mathbf{Y}) \in \mathbb{V}} {\lVert \mathbf{F}^{(e)}(\mathbf{X}) - \mathbf{Y} \rVert}_1$
    \If{$v^{(e)} < v$}
        \State Update running minimum validation metric: $v \gets v^{(e)}$
        \State Update running best weights and biases: $\boldsymbol{\Theta} \gets \boldsymbol{\theta}^{(e)}$
    \EndIf
\EndFor
\State \Return Write the trained NN weights and biases to disk: save($\boldsymbol{\Theta}$)
\end{algorithmic}
\end{algorithm}

\section{Estimating time decay constants}

Although the MWF parameter $c_1$ was of central interest in this paper, the estimation of the time decay constants $T_{21}$ and $T_{22}$ is also an independently important problem in magnetic resonance. In Tables \ref{tab:t21_parameter_estimation_results} and \ref{tab:t22_parameter_estimation_results} we include results for estimating $T_{21}$ and $T_{22}$ respectively, noting that our ILR networks $\textsc{(ND, Reg)}_{\textsc{NN}}$ and $\textsc{(ND, Reg)}_{\textsc{GCV}}$ output all three parameters at once. Therefore, no additional training or design is required to infer $T_{21}$ and $T_{22}$. 

\begin{table}[h]
    \centering
    \begin{tabular}{@{\extracolsep\fill}llll@{\extracolsep\fill}}
        \toprule
        \textbf{$\text{RMSE}_{T_{2,1}}$/SNR} & \textbf{5.0} & \textbf{50.0} & \textbf{100.0} \\
        \midrule 
        
        \textbf{(ND, Reg)$_{NN}$} & 14.45$^*$ & 10.41$^*$ & 8.55$^*$ \\
        \textbf{(ND, Reg)$_{GCV}$} & 14.45$^*$ & 10.42 & 8.63 \\
        \textbf{(ND, ND)} & 14.46 & 11.03 & 8.99 \\
        \textbf{TR-NLLS}, $\lambda_{oracle}$ & 23.03 & 18.46 & 18.40 \\
        \textbf{TR-NLLS}, $\lambda_{GCV}$ & 92.39 & 66.40 & 54.79 \\
        \textbf{NLLS} & $38.18 \times 10^{6}$  & $41.32 \times 10^{5}$ & $16.71 \times 10^{5}$ \\
        \bottomrule 
    \end{tabular}
    \caption{Root mean squared error (RMSE) in $T_{2,1}$ for each parameter estimation method. * indicates superior performance for the indicated SNR.}
    \label{tab:t21_parameter_estimation_results}
\end{table}

\begin{table}[h]
    \centering
    \begin{tabular}{@{\extracolsep\fill}llll@{\extracolsep\fill}}
        \toprule
        \textbf{$\text{RMSE}_{T_{2,2}}$/SNR} & \textbf{5.0} & \textbf{50.0} & \textbf{100.0} \\
        \midrule 
        \textbf{(ND, Reg)$_{NN}$} & 36.77$^*$ & 8.49$^*$ & 5.31$^*$ \\
        \textbf{(ND, Reg)$_{GCV}$} & 36.80 & 8.50 & 5.31$^*$ \\
        \textbf{(ND, ND)} & 36.86 & 8.57 & 5.35 \\
        \textbf{TR-NLLS}, $\lambda_{oracle}$ & 32.85 &  7.99 & 7.53 \\
        \textbf{TR-NLLS}, $\lambda_{GCV}$ & $15.29 \times 10^{3}$ & $59.20 \times 10^{2}$ & $53.68 \times 10^{2}$ \\
        \textbf{NLLS} & $41.61 \times 10^{7}$ & $12.86 \times 10^7$ & $13.69 \times 10^7$ \\
        \bottomrule 
    \end{tabular}
    \caption{Root mean squared error (RMSE) in $T_{2,2}$ for each parameter estimation method. * indicates superior performance for the indicated SNR.}
    \label{tab:t22_parameter_estimation_results}
\end{table}

\section{Disaggregating Table 2}

\begin{figure*}
    \centering
    \includegraphics[width=\linewidth]{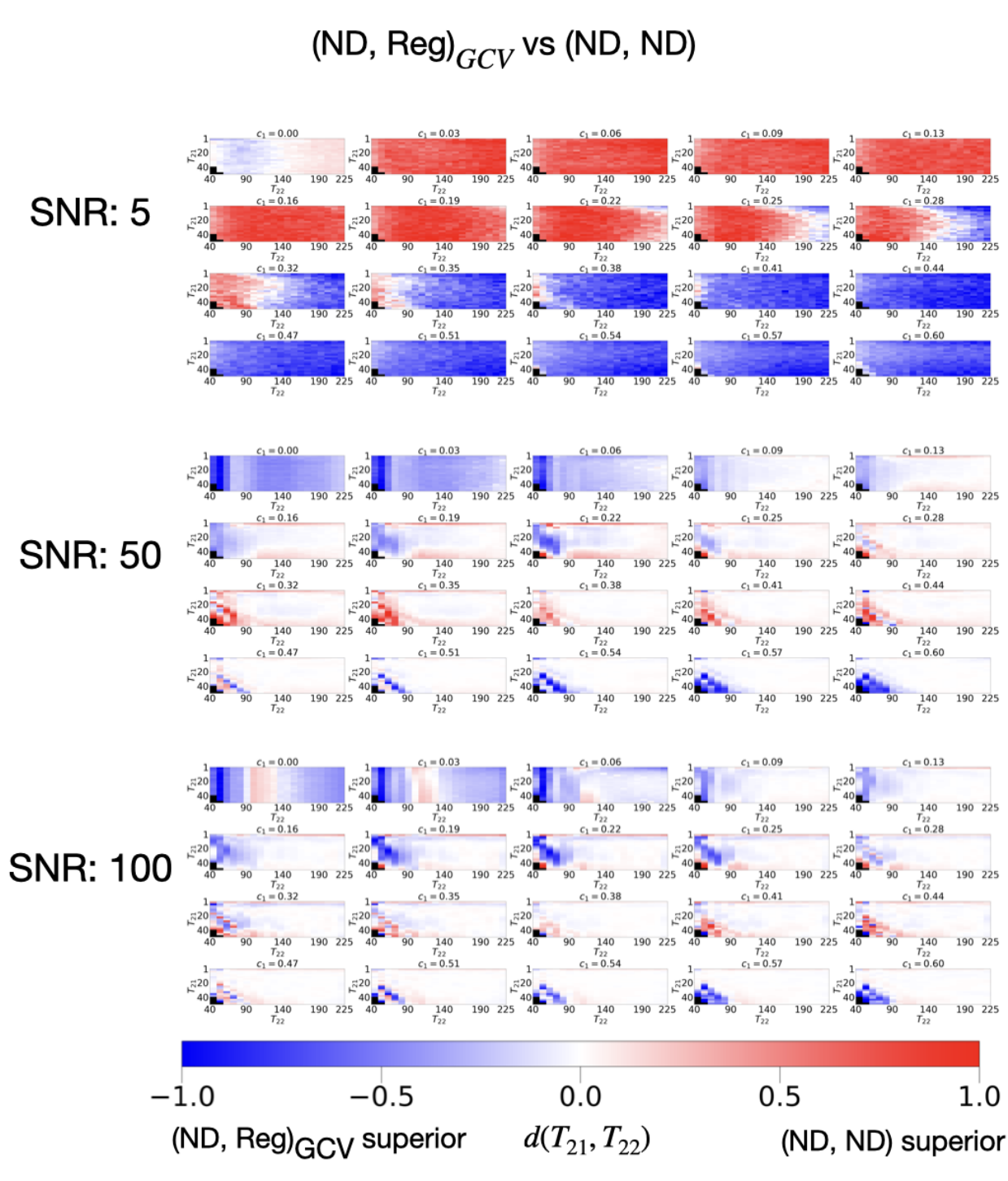}
    \caption{Top to bottom: Computing the differences in relative RMSE between $\textsc{(ND, Reg)}_{\textsc{GCV}}$ and $\textsc{(ND, ND)}$. Each panel is an image over the $(T_{21}, T_{22})$ space where the color blue (resp. red) represents the superiority of $\textsc{(ND, Reg)}_{\textsc{GCV}}$ (resp. $\textsc{(ND, ND)}$) in estimating the MWF value given by $c_{1}$ denoted above the panel.}
    \label{fig: NDRegGCVvsNDNDN}
\end{figure*}

\begin{figure*}
    \centering
    \includegraphics[width=\linewidth]{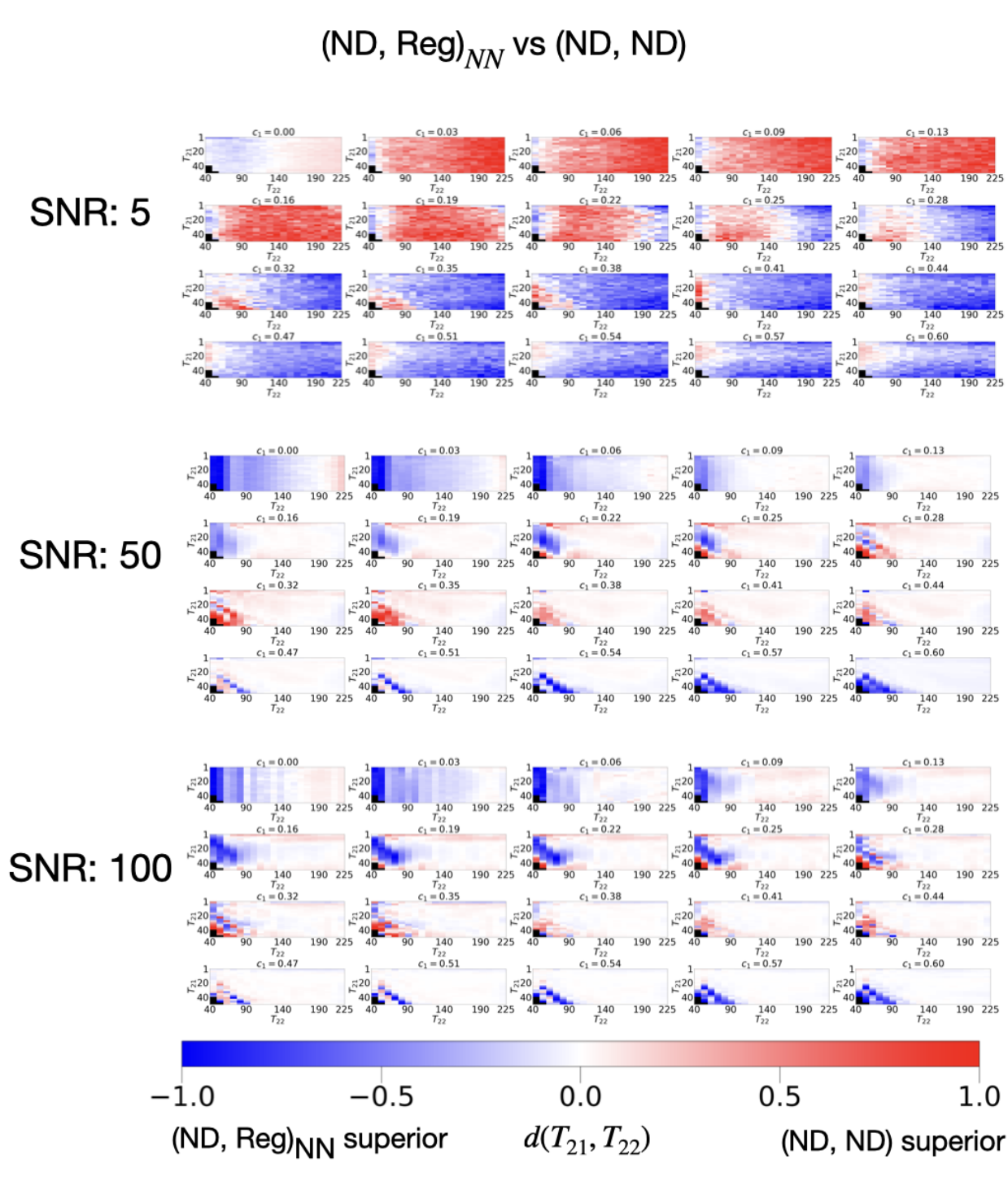}
    \caption{Results as for Figure \ref{fig: NDRegGCVvsNDNDN}, but for $\textsc{(ND, Reg)}_{\textsc{NN}}$.}
    \label{fig: NDRegNNvsNDNDN}
\end{figure*}

Although Table \ref{tab:c1_parameter_estimation_results} shows that our method outperforms both plain deep learning and classical methods overall, it is of interest to define the regions of parameter space $(c_1, T_{21}, T_{22})$ in which ILR is particularly beneficial over the plain neural network $\textsc{(ND, ND)}$. To investigate this issue, we compared the performance of the two NNs, $\textsc{(ND, Reg)}_{NN}$ and $\textsc{(ND, Reg)}_{GCV}$ against $\textsc{(ND, ND)}$ as follows. First, let $c^{(ND,Reg)}_{1}$, $c^{(ND,ND)}_{1}$ be the $c_1$ parameter estimates produced by the $\textsc{(ND, Reg)}$ and $\textsc{(ND, ND)}$ networks, respectively, where the subscript $NN$ or $GCV$ has been omitted for clarity here and below. Now given a fixed $c^{true}_1$, for each underlying pair  $(T_{21}, T_{22})$ we calculate the differences between the relative RMSE of the two estimates:
\begin{align}
    \widehat{d}(T_{21}, T_{22}) &= RMSE(c^{(ND,Reg)}_{1}, c^{true}_1)/c^{true}_1 - \\
    & RMSE(c^{(ND,ND)}_{1}, c^{true}_1)/c^{true}_1, \label{eq: NDReg v NDND} \\
    d(T_{21}, T_{22}) &= \frac{\widehat{d}(T_{21}, T_{22})}{\underset{T_{21}, T_{22}}{\textsf{max}}|\widehat{d}(T_{21}, T_{22})|}.\label{eq: NDReg v NDND normalized}
\end{align}
 Here the $RMSE$ is taken over 400 noise realizations present in the test set. From this, $\widehat{d}(T_{21}, T_{22}) < 0$ corresponds to $\textsc{(ND,Reg)}$ outperforming $(ND,ND)$ for the corresponding pair of $(T_{21}, T_{22})$. We also consider a normalized version of this error metric given by $d(T_{21}, T_{22})$ described in \eqref{eq: NDReg v NDND normalized} where the values of $\widehat{d}$ are rescaled to be in the interval $[-1,1]$. We visualize $d(T_{21}, T_{22})$ in Figures \ref{fig: NDRegGCVvsNDNDN} and \ref{fig: NDRegNNvsNDNDN}. Our findings are as follows.

 \emph{(Low SNR)} In the low SNR (SNR = $5.0$) regime (first panels from top in Figures \ref{fig: NDRegGCVvsNDNDN} and \ref{fig: NDRegNNvsNDNDN}), the two $\textsc{(ND, Reg)}$ networks markedly exceed the $\textsc{(ND, ND)}$ networks in performance for the $c_1 \geq 0.32$ region while the $\textsc{(ND, ND)}$ networks attain superiority when $c_1 \leq 0.32$. In this latter range of $c_1$ values, $\textsc{(ND, Reg)}_{\textsc{NN}}$ and $\textsc{(ND, Reg)}_{\textsc{GCV}}$ perform slightly differently. Specifically, when $0.25 \leq c_1 \leq 0.32$, $\textsc{(ND, Reg)}_{\textsc{NN}}$ outperforms $\textsc{(ND, ND)}$ for more values of $(T_{21}, T_{22})$ than $\textsc{(ND, Reg)}_{\textsc{NN}}$. 
 
 Thus, in regions of parameter space where $\textsc{(ND, Reg)}$ outperforms $\textsc{(ND, ND)}$, the ILR network with GCV selection given by $\textsc{(ND, Reg)}_{\textsc{GCV}}$ attains greater accuracy than $\textsc{(ND, Reg)}_{\textsc{NN}}$ (thus registering a lower aggregate RMSE in row 1 of Table \ref{tab:c1_parameter_estimation_results}). However, $\textsc{(ND, Reg)}_{\textsc{NN}}$ outperforms $\textsc{(ND, ND)}$ for larger regions of parameter space than $\textsc{(ND, Reg)}_{\textsc{GCV}}$.

 \emph{Medium and high SNR} When $SNR = 50, 100$, both of the $\textsc{(ND, Reg)}$ networks exhibit distinct patterns in three different ranges of $c_1$ space: 

 \begin{enumerate}
     \item $0.0 \leq c_1 \leq 0.25$: In the low $c_1$ range the $\textsc{(ND, Reg)}$ networks outperform $\textsc{(ND, ND)}$. This gain in performance is specifically more pronounced in the $T_{22} - T_{21} \approx 20$ region, as evidenced by the diagonal blue bands in the lower left corners in the central cells of the SNR 50 and 100 panels in Figures \ref{fig: NDRegGCVvsNDNDN} and \ref{fig: NDRegNNvsNDNDN}. However, as the difference $T_{22} - T_{21}$ decreases further to zero, there is a greater prevalence of red pixels in the left corner of each cell, indicating that when the signal is nearly monoexponential, ILR likely introduces inaccuracies stemming from model misspecification. More generally, the difference between the decay times, $T_{22} - T_{21},$ is an indicator of the ill-posedness of the biexponential problem. Our results in Figures \ref{fig: NDRegGCVvsNDNDN} and \ref{fig: NDRegNNvsNDNDN} for this range of $c_1$ space may thus be qualitatively described as follows: in the regions of parameter space with a large weighting of the faster decay (i.e $c_1 \leq 0.25$), ILR is particularly advantageous when the problem features an intermediate level of ill-posedness (i.e., $T_{22} - T_{21} \approx 20$). However, when the time constants are nearly identical, $\textsc{(ND, ND)}$ exhibits superior performance.
     \item $0.25 \leq c_1 \leq 0.4$: In the intermediate $c_1$ range, the $\textsc{(ND, ND)}$ networks largely outperform the $\textsc{(ND, Reg)}$ networks. Moreover, the $\textsc{(ND, ND)}$ networks specifically attain lower RMSE in the ill-posed region where $T_{21} \approx T_{22}$. 
     \item $0.4 \leq c_1 \leq 0.6$: This is the region of parameter space where the two time-decays are nearly equally weighted. In this region of equal weighting, ILR is substantially better than plain neural network estimation, with the most significant improvements occurring in the ill-posed regions. This phenomenon persists for both medium and high SNRs for both $\textsc{(ND, Reg)}$ networks. 
 \end{enumerate}

\emph{Does ILR behave like classical regularization?} For the NLLS estimator, regularization improves parameter estimation in the regions of ill-posedness, that is, when  $T_{21}$ and $T_{22}$ are within, roughly, a factor of 2 or 3 of each other. A natural question then is whether ILR acts as a regularizer for the parameter estimation problem; this can be evaluated by examining its performance in ill-posed and well-posed regimes of parameter space. In Figures \ref{fig: NDRegGCVvsNDNDN}and \ref{fig: NDRegNNvsNDNDN}, the lower left corners are the most ill-posed regimes. As seen, the effect of ILR is to improve overall performance across the diagrams, but without the improvement generally being concentrated in the ill-posed regime.

\section{Brain data}

\subsection{MRI acquisition protocol}

After obtaining informed consent, imaging was performed with a 3T Philips MRI system (Achieva, Best, The Netherlands), with an internal quadrature bodycoil for transmission and an eight-channel phased array head coil for reception. We used the 3D GRASE (gradient and spin-echo) sequence to acquire data from the brain of a healthy 41-year-old male. Echo data were acquired for 32 echoes, with acquisition times of $t_n=n \times \Delta t$ where the echo spacing $\Delta t=11.3ms$. An echo planar imaging (EPI) acceleration factor of 3 was used, with a field of view (FOV) $278 mm \times 200 mm \times 30 mm$, over an acquisition matrix size of $185 \times 133 \times 10$, for a voxel size of $1.5 mm \times 1.5 mm \times 1.5 mm$, with $TR=1000ms$. The image matrix was zero filled in k-space to yield a $1 mm \times 1 mm \times 1 mm$ reconstruction. The duration of the scan was approximately $10$ minutes. Sampling was performed at echo peaks and formed the signal data for analysis. 

\subsection{Noise filtering and quantification}

\begin{figure}[h]
    \centering
    \includegraphics[width=0.6\textwidth]{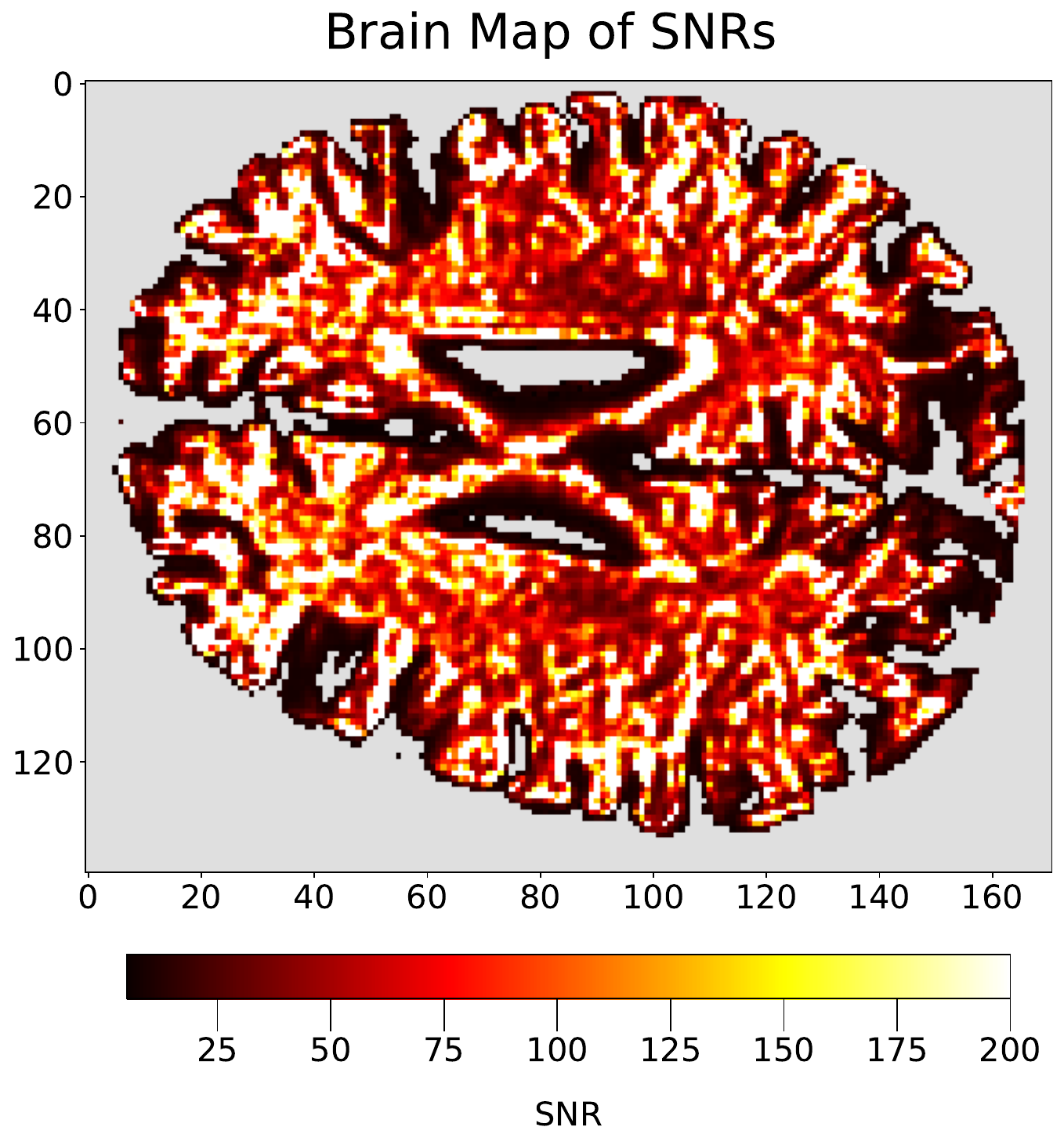}
    \caption{The SNR for each voxel was calculated by dividing the signal amplitude at the first echo time by the signal floor (offset value) as estimated with NLLS: $\frac{SA_{TE_0}}{offset /\sqrt{\pi / 2}}$. Excluded voxels are colored grey.}
    \label{fig: brain SNR}
\end{figure}

Lacking a true "ground truth" for in vivo data, we used the NESMA filter \cite{bouhrara2018use} to obtain high SNR images from which to extract parameter values that serve as our (necessarily flawed) standard for comparison. This also provided a base image to which we added noise to achieve the levels of SNR as described in the main text, allowing us to implement a targeted application of an ILR network trained on data of low, medium, or high SNR. Figure \ref{fig: brain SNR} displays the calculated SNRs.  

\subsection{Akaike information criterion (AIC)}

We wish to apply regularization only to signals well-described as biexponential rather than monoexponential. In the latter case, our biexponential model is undetermined, and the resulting parameter variances cannot be remediated by regularization or any other numerical approach. We used the Akaike information criterion (AIC) to identify biexponential signals. The AIC provides a criterion for determining whether the improved numerical fit that inevitably results from an increase in parameters for nested models is statistically meaningful. In the context of least squares regression, the AIC can be derived from the residual sum of squares (RSS). For a linear regression model with normally distributed errors, the likelihood \( L \) is proportional to \( \exp\left(-\frac{RSS}{2\sigma^2}\right) \). This leads to the following expression for the AIC:
\[
AIC = n \log(\text{RSS}/n) + 2k
\]
Here \( n \) is the number of data points, \( \text{RSS} \) is the residual sum of squares, and \( k \) is the number of parameters. In the context of MWF imaging, we consider two competing models for explaining the data: 
\begin{align}
    \textbf{G}_{monoexp}(t, c_1, T_{21}, b) &= b + c_1 \exp{(-t/T_{21})}, \\
    \textbf{G}_{biexp}(t, c_1, T_{21}, c_{2}, T_{22}, b) &= b + c_1 \exp{(-t/T_{21})} \\
    &+ c_2\exp{(-t/T_{22})}.
\end{align}
Here $b$ is an offset, $c_1$, $c_2$ are compartment sizes, and $T_{21}$ and $T_{22}$ are exponential time constants, as defined in the text.
The AIC provides a method to compare the suitability of these models. In the context of MWF imaging each voxel can be regarded as containing a mono or bi-exponentially decaying signal corrupted by Rician noise. The RSS is not strictly an estimate of the likelihood of the data for non-Gaussian noise, but our SNR range after NESMA filtering is sufficiently large that the difference between Gaussian and Rician noise can be neglected.

\nocite{*}

\clearpage
 \bibliographystyle{elsarticle-num} 
 \bibliography{references}



\end{document}